\documentclass[journal,twoside]{IEEEtran}
		\usepackage[utf8]{inputenc}
		\usepackage{cite}
		\usepackage{graphicx}
		\usepackage{tikz}
			\usetikzlibrary{shapes.geometric}
			\usetikzlibrary{calc}
		\usepackage{tikz-3dplot}
		\usepackage{datatool}
		\usepackage{pgfplots,pgffor}
			\pgfplotsset{compat=newest,every axis/.append style={font=\small},}
		\usepackage{amsmath,amssymb,amsthm}
		\usepackage{url}
		\usepackage{etoolbox}
		\usepackage{xpatch}
		\usepackage{algorithmic}
			\makeatletter
			\xpatchcmd{\algorithmic}{\itemsep\z@}{\itemsep=8pt plus4pt}{}{}
			\makeatother
		\usepackage{array}
		\usepackage[caption=false,font=normalsize,labelfont=sf,textfont=sf]{subfig}
	
		\usepackage{subfiles}
		\usepackage{hyperref}
		
		\IfFileExists{macros.tex}{	
	\def\natu{\mathbb{N}}
	\def\reals{\mathbb{R}}
	\def\integ{\mathbb{Z}}
	
	\newcommand{\supp}[1]{\operatorname{supp}\left( #1 \right)}
	\renewcommand{\complement}[1]{#1^{\mathsf{c}}}
	
	\def\sigmax{\sigma_{\max}}
	\def\dsigma{\tilde{\sigma}}
	\def\dsigmax{\tilde{\sigma}_{\max}}
	\def\ka{\kappa_\mathrm{a}}
	\def\kd{\kappa_\mathrm{d}}
	\def\pos{\mathbf{r}}
	\def\ae{\mathrm{a.e.}}
	\def\obs{d_{\mathrm{obs}}}
	\def\dobs{\tilde{d}_{\mathrm{obs}}}
	\def\dobsq{\tilde{d}_{\mathrm{obs},q_2}}

	\def\wninf{\left\| w \right\|_{\mathrm{L}^\infty\left(\reals^2\right)}}
	\def\dint{\mathrm{d}}
	\newcommand{\Restr}[1]{\operatorname{R}_{#1}}
	\newcommand{\Exten}[1]{\operatorname{E}_{#1}}
	
	\newcommand{\listint}[1]{\lbrace 1,2,\dots,#1\rbrace}
	\newcommand{\matrices}[1]{\operatorname{\mathbb{T}}\left( #1 \right)}
	\newcommand{\matricesP}[1]{\operatorname{\mathbb{T}_+}\left( #1 \right)}
		\def\DenSpace{\mathcal{D}}
		\newcommand{\prodDen}[2]{ \left( #1|#2 \right)_{\DenSpace} }
		\newcommand{\normDen}[1]{ \left\|#1\right\|_{\DenSpace} }
		\def\ASpace{\mathcal{A}}
		\newcommand{\prodA}[2]{ \left( #1|#2 \right)_{\ASpace} }
		\newcommand{\normA}[1]{ \left\|#1\right\|_{\ASpace} }
		\newcommand{\Leb}[2]{\mathrm{L}^#1\left( #2 \right)}
		\newcommand{\LebTwo}[1]{\Leb{2}{#1}}
		\newcommand{\LebOne}[1]{\Leb{1}{#1}}
		\newcommand{\normLebTwo}[2]{ \left\| #2 \right\|_{\LebTwo{#1}} }
		\newcommand{\normLebOne}[2]{ \left\| #2 \right\|_{\LebOne{#1}} }
		\newcommand{\prodLebTwo}[3]{ \left( #2|#3 \right)_{\LebTwo{#1}} }
		
		\newcommand{\opers}[2]{ \mathcal{L}\left(#1,#2\right) }
		\newcommand{\NullSpace}[1]{\operatorname{\mathcal{N}}\left(#1\right)}
		\def\X{\mathcal{X}}

		\newcommand{\discreg}[1]{{\Lambda}_{\scriptstyle #1}}
	\newcommand{\changetoAlg}{ 
		\renewcommand{\figurename}{Alg.}
	}
	\newcommand{\changetoFig}{
		\renewcommand{\figurename}{Fig.}
	}
	
	\makeatletter%
	\@ifclassloaded{beamer}%
		{}%
		{
		\newtheorem{definition}{Definition}
		\newtheorem{lemma}{Lemma}
		\newtheorem{property}{Property}
		\newtheorem{theorem}{Theorem}
		
		}
		\makeatother%

		\def\figs{figs}
		\def\secs{secs}
		\def\bib{.}
	
	}{}
		
\begin{document}

			\title{Cell Detection by Functional Inverse Diffusion and Non-negative Group Sparsity---Part I: Modeling and Inverse Problems}
			\author{
				\IEEEauthorblockN{Pol del Aguila Pla, \emph{Student Member, IEEE}, and Joakim Jaldén, \emph{Senior Member, IEEE}\thanks{}\thanks{
				Manuscript received September 21, 2017; revised March 31, 2018 and July
				9, 2018; accepted August 24, 2018. Date of publication September 3, 2018;
				date of current version September 14, 2018. The associate editor coordinating
				the review of this manuscript and approving it for publication was Prof. Mark
				A. Davenport. This work was supported in part by Mabtech AB and in part by
				the Swedish Research Council (VR) under Grant 2015-04026. \emph{(Corresponding
				author: Pol del Aguila Pla.)} }\thanks{
				The authors are with the Department of Information Science and Engineering,
				School of Electrical Engineering and Computer Science, KTH Royal
				Institute of Technology, Stockholm 11428, Sweden (e-mail:, poldap@kth.se;
				jalden@kth.se).} \thanks{
				This paper has supplementary downloadable material available at http://
				ieeexplore.ieee.org, provided by the authors. The material includes detailed
				derivations of some key steps and further experimental results. This material is
				$533$ kB in size.} \thanks{
				Color versions of one or more of the figures in this paper are available online
				at http://ieeexplore.ieee.org.}\thanks{
				\textbf{Author's own archival version.} Digital Object Identifier of the original manuscript: 10.1109/TSP.2018.2868258.}}
				}
		\makeatletter
		\ifCLASSOPTIONpeerreview
			\markboth{IEEE TRANSACTIONS ON SIGNAL PROCESSING, VOL. 66, NO. 20, OCTOBER 15, 2018 
			}%
			{\MakeUppercase{\@title}}
		\else
			\markboth{IEEE TRANSACTIONS ON SIGNAL PROCESSING, VOL. 66, NO. 20, OCTOBER 15, 2018 
			}%
			{DEL AGUILA PLA AND JALD\'EN: CELL DETECTION BY FUNCTIONAL INVERSE DIFFUSION AND NON-NEGATIVE GROUP SPARSITY---PART I}
		\fi
		\makeatother
		\IEEEpubid{\begin{minipage}{\textwidth}
			    \vspace{20pt}
			    \begin{center}
			      1053-587X~\copyright~2018 IEEE. Translations and content mining are permitted for academic research only. 
			      Personal use is also permitted, but republication/redistribution requires IEEE permission. 
			      See http://www.ieee.org/publications standards/publications/rights/index.html for more information.
			    \end{center}
			   \end{minipage}
		}
	
	\maketitle
	\begin{abstract}
		In this two-part paper, we present a novel framework and methodology to 
analyze data from certain image-based biochemical assays, e.g., ELISPOT 
and Fluorospot assays. In this first part, we start by presenting a 
physical partial differential equations (PDE) model up to image acquisition for these 
biochemical assays. Then, we use the PDEs' Green function to derive a
novel parametrization of the acquired images.
This parametrization allows us to propose a functional
optimization problem to address inverse diffusion. 
In particular, we propose a non-negative group-sparsity regularized optimization problem with the goal
of localizing and characterizing the biological cells involved in the said assays. 
We continue by proposing a suitable discretization scheme that enables both the generation of synthetic
data and implementable algorithms to address inverse diffusion.
We end Part I by providing a preliminary comparison between the results of our methodology and an expert human labeler 
on real data. Part II is devoted to providing an accelerated proximal gradient 
algorithm to solve the proposed problem and to the empirical validation of our methodology.
	\end{abstract}
	\begin{IEEEkeywords}
		Inverse problems, Biomedical imaging, Convex optimization, Source localization, Biological modeling
	\end{IEEEkeywords}

		\ifCLASSOPTIONpeerreview
		\begin{center} 
			\bfseries EDICS Category: 3-BBND 
		\end{center}
		\fi
		\IEEEpeerreviewmaketitle
	
			\providebool{tot}
			\booltrue{tot}

	
		\section{Introduction} \label{sec:Intro}

	\IEEEPARstart{B}{iological} processes in which cells generate particles that diffuse in a solution and bind to
	receptors are ubiquitous 
	\cite{Lagerholm1998,Lieto2003,Berezhkovskii2004,Plante2011,Karulin2012,Plante2013}. 
	Such processes are often measured using biochemical assays where cells are contained in a well with a 
	receptor-coated bottom, and an image of the resulting density of bound particles is obtained. Examples include
	the ELISPOT \cite{Czerkinsky1983} and Fluorospot \cite{Gazagne2003} assays.
	If particles bind relatively close to their origin, the cells that generated these particles (active cells)
	can be localized in the obtained image. Localization enables counting, and therefore, quantitative
	studies of the proportions of active cells within the cell population under study. 
	Thereby, these assays provide answers to relevant questions in fields ranging from biochemical, 
	pharmacological, and medical research \cite{Karulin2012,Dillenbeck2014,Martinez-Murillo2016}, 
	to the diagnosis of specific diseases \cite{Meier2005,2015}. 
	Hence, source localization (SL) algorithms are critical to the 
	development of automated analysis systems for high-throughput
	pharmacological and medical applications.
	In this first part of our paper, we present a $2$-dimensional (2D) equivalent diffusion model for the density of bound 
	particles generated by a 3D reaction-diffusion-adsorption-desorption process. We then propose a functional 
	optimization framework for inverse 2D diffusion that promotes stationary-source explanations of the observed data.
	Then, we present a discretization scheme that allows both for the synthesis of realistic data and for numerical 
	solutions to the proposed optimization problem.  
	Part II of this paper \cite{AguilaPla2017a} is devoted to algorithmic solutions to solve
	this optimization problem.
	
	\IEEEpubidadjcol
	The accuracy of SL algorithms becomes critical when characterizing cell sub-populations by multiplex assays, e.g. 
	Fluorospot \cite{Gazagne2003}.
	Multiplex assays allow different kinds of particles to be independently and simultaneously measured, yielding 
	co-located images. The results of their analyses are then merged to detect which cells were producing which 
	combinations
	of particle types. This data fusion is conducted based on the only comparable feature of multiple-secreting cells 
	in each of the images, i.e., their location. Therefore, localization accuracy has a direct impact on the 
	estimated proportions, i.e., on the accuracy of multiplex assays.
	The optimization framework we propose uses a non-parametric model-based approach to produce results 
	that enable accurate SL and, thereby, accurate results in multiplex assays. 
	In finalizing this first part, we provide results on real data by comparing our solution to the labeling of a 
	 human expert. In Part~II of this paper \cite{AguilaPla2017a}, we provide a thorough evaluation of the 
	proposed methodology using synthetic data.

	SL on 2D or 3D data from linear observation models
	has been widely studied for biologic 
	\cite{Olivo-Marin2002,Rebhahn2008,Pan2010a,Smal2010,Kimori2010,Ram2012,Zhao2014,Basset2015,Kervrann2016},
	astronomic 
	\cite{Starck2002,Giovannelli2005}, 
	acoustic 
	\cite{Ehrenfried2006,Markovic2015},
	heat conduction 
	\cite{Ternat2012,Zhang2016},
	and environmental applications 
	\cite{Matthes2005,Hamdi2007,Hamdi2012},
	as well as in more generic settings 
	\cite{Kaaresen1997,Li2004,Mazet2004,Kail2012,Selesnick2014}.
	Parametric approaches to SL have been thoroughly investigated when the source-map is 
	observed through a convolutional operator 
	\cite{Li2004,Mazet2004,Ehrenfried2006,Zhao2014}. 
	In particular, sparsity-based approaches have been shown to have many favorable
	properties in this case (see \cite{Duval2015} and references therein). 
	To our knowledge, SL from data obtained from linear diffusion 
	has only been addressed parametrically 
	\cite{Matthes2005,Hamdi2007,Hamdi2012}. 
	A downside of parametric approaches is that the full characterization of the observation system is 
	seldom available and, thus, it has to be specifically measured \cite{Matthes2005} or estimated
	\cite{Schwab2012}. This implies additional costs for practical use, which hinder scalability.
	Non-parametric approaches to image-based SL can be divided in two categories. On one hand,
	model-independent approaches work solely on image properties, yielding heuristic methods to find dot-like 
	shapes in images \cite{Olivo-Marin2002,Rebhahn2008,Pan2010a,Smal2010,Kimori2010,Ram2012}. These are combined with 
	generic data-analytic procedures to address measurement-noise and yield results that may be satisfactory, but 
	are biased by the arbitrary heuristics and tend to over- or under-react to small perturbations. 
	On the other hand, model-based approaches use the structure of the problem, 
	exploiting properties specific to the process that generated the data without requiring previous 
	measurement of the intrinsic values that regulate it. Most representative of these 
	model-based non-parametric approaches are blind deconvolution methods, e.g., \cite{Kail2012}.
	The inverse diffusion approach we present is model-based and non-parametric, providing a robust and 
	scalable methodology to address SL in reaction-diffusion-adsorption-desorption models.
		
	The inversion of diffusion equations has been widely studied \cite{Zhao2011,Hon2011,Wen2013,Zhang2016}. However,
	most methods address the ill-posedness of the problem by regularizing it to favor smooth solutions,
	i.e., by aiming to provide the least sharp release of particles over time and space that explains the data.
	In SL, however, one is assuming that the data has been created by localized sources. Consequently, one would want to
	favor the most localized, i.e., spatially sharpest, release of particles over time that explains the data. There 
	are some approaches
	that target diffusion-based SL, or, closely related, the recovery of non-smooth solutions from inverse diffusion 
	problems 
	\cite{Matthes2005,Hamdi2007,Hamdi2012,Ternat2012}. 
	However, inverse diffusion leads to very different problem formulations depending
	on the restrictions one imposes on the generation of particles, the boundary conditions of the medium, i.e., the 
	additional effects one takes into account (such as adsorption and desorption), and the kind of measurements one has 
	access to.
	\cite{Matthes2005,Hamdi2007,Hamdi2012} study 2D reaction-advection-diffusion, but only 
	consider particles released from a single point, and intend to localize it as accurately as possible. 
	This allows for a study specific to SL, in the sense that generic inversion of the diffusion equation is unnecessary.
	In particular, \cite{Hamdi2007,Hamdi2012} provide technical results on the identifiability of 
	a single source. In contrast, \cite{Ternat2012} studies 1D diffusion with known-concentration boundary conditions, 
	and while
	it allows for an initial concentration of particles that varies throughout the considered area, it does not contemplate 
	the
	effect of the continuous generation of particles (reaction). In this paper, we study 3D 
	reaction-diffusion-adsorption-desorption,
	we do not impose any restrictions on reaction, and we do not presume any artificial Dirac behavior in the 
	spatial or temporal domains. Instead, we use regularization to favor explanations of the data that are spatially 
	sparse and temporally continuous, as stationary cells releasing particles would be.

	\subsection{Notation} \label{ssec:Notation}
		
		When sets and spaces of numbers are involved, we will use either standard notation such as
		$\reals_+=\left[ 0, +\infty\right)$, $\bar{\reals}=\reals \cup \lbrace -\infty, +\infty \rbrace$ 
		and $\bar{\reals}_+=[0,+\infty]$
		or capital non-Latin letters, e.g., we will use $\Omega = \reals^2 \times \reals_+$ because of the many times
		we will refer to functions in this particular support. When discussing locations in $\reals^2$,
		we will note them as bold face letters, e.g., $\pos \in \reals^2$.
		
		When discussing functional sets and spaces, we will use capital calligraphic notation, such as $\X$ for a generic 
		normed space,
		which will have norm $\|\cdot\|_{\X}$. If $\X$ is also a Hilbert space, $\X$ will have scalar product 
		$\left( \cdot | \cdot \right)_{\X}$.
		For any functional space $\X$, $\X_+\subset \X$ is the cone of non-negative
		functionals. Specifically, if $\X$ contains functionals $f:\mathcal{Y}\rightarrow \reals$, then
		$
			\X_+=\left\lbrace f\in \X : f(y) \geq 0, \forall y \in \mathcal{Y} \right\rbrace \subset \X 
		$.
		For any set $\mathcal{Z} \subseteq \X$, its $(\infty,0)$-indicator function is the function $
		\delta_\mathcal{Z}:\X \rightarrow \lbrace 0,+\infty \rbrace$
		such that $\delta_\mathcal{Z}(x) = 0$ if $x \in \mathcal{Z}$ and $\delta_\mathcal{Z}(x) = +\infty$ if
		$x \in \complement{\mathcal{Z}}=\X\setminus\mathcal{Z}$, while 
		its $(0,1)$-indicator function is the function 
		$i_\mathcal{Z}:\X \rightarrow \lbrace 0,1 \rbrace$ such that $i_\mathcal{Z}(x) = 1$  if $x \in \mathcal{Z}$ and 
		$i_\mathcal{Z}(x) = 0$ if $x \in \complement{\mathcal{Z}}$.
		
		When discussing a specific functional $f\in \X$, $f_+:\mathcal{Y} \rightarrow \reals$ will be its positive part,
		i.e., 
			$	f_+(y) = \max\lbrace f(y), 0 \rbrace\,,\forall y \in \mathcal{Y}$.
		The support of the functional $f\in\X$ will be written as
			$\supp{f} = \lbrace y\in\mathcal{Y}: f(y) \neq 0 \rbrace\subset \mathcal{Y}$.
		Finally, for any two given functions 
		$f,g:\reals^N\rightarrow \reals$ for some $N\in\natu$, we refer to their convolution as $(f * g)$ and to the $j$-th
		convolutional power of $f$ as $f^{j*}$.
		
		When discussing operators, if $\mathcal{Z}$ is some normed space, we will write $\opers{\X}{\mathcal{Z}}$ for 
		the space of linear continuous operators
		from $\X$ to $\mathcal{Z}$. Coherently with the notation above, this space of operators will have norm
		$\|\cdot\|_{\opers{\X}{\mathcal{Z}}}$. We will note operators as $A$ or $B$, e.g., 
		$ B \in \opers{\X}{\mathcal{Z}}$.
		For any such $B$, we will refer to its adjoint as $B^*\in \opers{\mathcal{Z}}{\X}$. Recall that,
		if $\X$ and $\mathcal{Z}$ are Hilbert spaces, $\left(Bx|z\right)_{\mathcal{Z}}=(x|B^*z)_{\X}$,
		for any $x\in\X$ and any $z\in\mathcal{Z}$.
		
		When discussing matrices and tensors, the space of real $M$-by-$N$ matrices for some $M,N\in\natu$ is $\matrices{M,N}$, while its element-wise 
		positive cone is $\matricesP{M,N}$. For a specific matrix $\tilde{f}\in\matrices{M,N}$, we specify it as a group of its elements 
		$\left\lgroup \tilde{f}_{m,n} \right\rgroup$ for $m\in\listint{M}$ and $n\in\listint{N}$. 
		For tensors, we work analogously by adding appropriate indexes, e.g., $\tilde{f}\in\matrices{M,N,K}$ and $\left\lgroup \tilde{f}_{m,n,k} \right\rgroup$
		for $k\in\listint{K}$.
		
		When presenting our statements, we will refer to them as properties if they are not novel, but are necessary 
		for clear exposition,
		lemmas if they contain minor novel contributions and theorems if they constitute major novel contributions.

	\ifbool{tot}{}{
		\bibliographystyle{IEEEtran}
		\bibliography{IEEEabrv,\bib/multi_deconv}
	}

		\section{Data Model} \label{sec:DataModel}

	\subsection{Physical model} \label{ssec:PhyMod}
	
\begin{figure}
			\centering
			\includegraphics[keepaspectratio=true,clip=true,trim=3in 8.9in 3.4in 1in,scale=1.63]{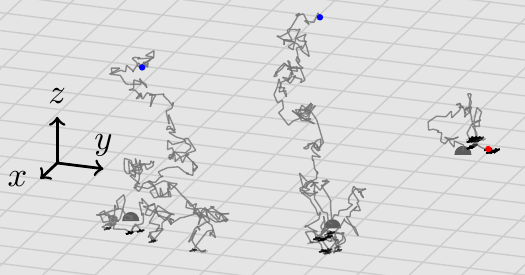}
			\\
			\footnotesize (a) Particles motion model \\ \vspace{4pt}
			\includegraphics[keepaspectratio=true,scale=1]{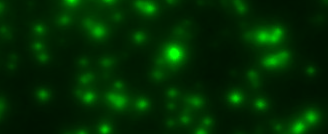} \\
			\footnotesize (b) Typical observation
			\caption{ 
					(a) Visualization, at a particle level, of the physical data model described in 
					Section~\ref{ssec:PhyMod}.  
					Three particles, each secreted by a different cell (dark gray) immobilized on the plane (light gray),
					follow a Brownian motion.
					When the particles hit the plane, they might bind to it (adsorption; black marks). After a time, they may 
					disassociate
					(desorption) and continue their Brownian motion. At the end of the experiment, i.e., at time $T$, 
					they may be free (blue dots) and thus 
					not imaged or bound to the surface (red dot) and thus contribute to the final image. Note that while
					the relative scale
					between movement and the pixel size of a potential imaging sensor is consistent with accurate 
					physical parameters, the relative 
					scale of cells and particles was selected for clear visualization.
					(b) Example section of an image observation from a Fluorospot assay. 
					Here, FITC dye was used as a marker for trapped IFN-$\gamma$ molecules, and the resulting 
					$512~\mathrm{nm}$ fluorescence was isolated by optic filters and subsequently captured by a 
					color camera at approximately $1$ to $1$ magnification.
					\label{fig:physics}
					} 
		\end{figure}

		We consider a physically motivated 3D stochastic model where cells are immobilized on a flat surface, represented
		here 
		by the $xy$-plane. Some of these cells are active, i.e., they release particles into a medium located above the surface, in the half-space $z\geq 0$. 
		Released particles then move in a 3D isotropic Brownian motion. 
		The same surface where the cells reside is evenly coated with imperfect receptors
		tuned specifically to the released particles. Therefore, particles diffusing in the medium that collide with the 
		surface may bind to it, but also, bound
		particles may disassociate from the surface after some time.
		Particles bound to the surface at a time $T$, i.e., when the experiment finishes, are then tagged with some 
		visible marker, and their density is imaged. 
		This produces spots around each active cell in the captured image. The model is illustrated at a particle level in 
		Fig.~\ref{fig:physics}, which also includes
		a section from a typical observation from a Fluorospot assay.
		Note here that cells are tens of $\mu\mathrm{m}$s in diameter and that the particles of interest are typically 
		of a few $\mathrm{nm}$s in diameter \cite{Reth2013}. 
		Moreover, the visible spots produced by active cells in these assays are typically no more than $200~\mu\mathrm{m}$ 
		in diameter. Because these assays are conducted inside wells of approximately $7\,\mathrm{mm}$ in diameter, we 
		disregard the effects of the borders of the well 
		for the rest of the paper.

		We assume that the medium is homogeneous and that the particle concentrations are low 
		enough so that we can consider
		the binding affinity and disassociation rate of the surface constant and uniform. These assumptions imply that 
		we can model the movement of 
		individual particles as independent of each other, which renders the model spatially invariant in any direction on 
		the $xy$-plane. These assumptions
		are also consistent with the models considered in, e.g., 
		\cite{Lagerholm1998,Lieto2003,Berezhkovskii2004,Plante2011,Karulin2012,Plante2013}.

		Consider the function 
			$c:\reals^2\times \reals_+ \times \reals_+ \rightarrow \reals_+$
		such that $c(x,y,z,t)~[\mathrm{m}^{-3}]$ is the time-varying concentration of free particles in the medium. This 
		concentration $c$ is modeled 
		via the 3D homogeneous diffusion equation,
		\begin{subequations} \label{eq:pde}
			\begin{IEEEeqnarray}{c} \label{eq:pde-diffusion}
				\frac{\partial}{\partial t} c = D \Delta c \, ,
			\end{IEEEeqnarray}
			where
				$\Delta = \partial^2/\partial x^2 +  \partial^2/\partial y^2 +  \partial^2/\partial z^2$
			is the Laplace operator and $D~[\mathrm{m}^{2}\mathrm{s}^{-1}]$ is the diffusion constant of the released 
			particles in the medium. 
			Consider now the function
				$d:\reals^2\times[0,+\infty) \rightarrow \reals_+$
			such that $d(x,y,t)~[\mathrm{m}^{-2}]$ is the surface density of bound particles at time $t$. This density 
			$d$ is coupled to $c$ via the adsorption-desorption 
			boundary condition \cite{Agmon1984}, given by
			\begin{IEEEeqnarray}{c} \label{eq:pde-surface}
				\frac{\partial}{\partial t} d = \kappa_\mathrm{a} c \big|_{z=0} - \kappa_\mathrm{d} d\,,
			\end{IEEEeqnarray}
			and via the condition on the flow of particles away from the surface \cite{Plante2011}, given by 
			\begin{IEEEeqnarray}{c} \label{eq:pde-boundaryflow}
				-D \frac{\partial}{\partial z} c \big|_{z=0} = s + \kappa_\mathrm{d} d - \kappa_\mathrm{a} c\big|_{z=0}\,.
			\end{IEEEeqnarray}
		\end{subequations}
		Here, the function 
			$s:\reals^2\times[0,+\infty) \rightarrow \reals_+$
		is such that $s(x,y,t)~[\mathrm{m}^{-2} \mathrm{s}^{-1}]$ denotes the source density rate (SDR) of new particles 
		released from cells residing 
		at the surface, and $\ka~[\mathrm{m}\mathrm{s}^{-1}]$ and $\kd~[\mathrm{s}^{-1}]$ are the adsorption and 
		desorption
		constants, respectively. We will assume here that $c(x,y,z,t) = 0$, $d(x,y,t) = 0$, and $s(x,y,t) = 0$ for 
		$t < 0$, i.e., that before
		starting the experiment no particles have been generated or are present.
	
	\subsection{Observation model} \label{sec:DataModel:ObsModel}
	
		Our primary interest in \eqref{eq:pde} lies in characterizing the surface density $d$ at the time $T$ at which 
		it is imaged, in terms of the SDR $s$.
		Therefore, we consider the concentration $c$ in the medium to be only an intermediate nuisance parameter. For 
		notational brevity we will write as $\pos = (x,y)$
		the spatial coordinates of a generic point on the surface $z=0$, and refer to the final image observation as 
		$\obs$, i.e., $\obs(\pos) = d(\pos,T)$.
		Note here that while $\obs$ is considered to be exactly equal to the density of particles bound to the surface, 
		in practice, imaging sensors will have different
		sensitivities and, thus, there will always be a factor of scale $\alpha>0$, which we will disregard in this 
		paper. Further limitations of imaging sensors,
		such as finite dimensionality and imperfections in the optical and electrical
		systems involved, are discussed in Section~\ref{ssec:SensLim}.
		
		To obtain a suitable characterization of the mapping from $s$ to $\obs$, we will follow the arguments given 
		in \cite{Berezhkovskii2004} and 
		interchangeably rely on macroscopic arguments pertaining to the evolution of particle distributions, governed 
		by \eqref{eq:pde}, and microscopic 
		arguments pertaining to the behavior of individual particles \cite{Ursell2007}. Note now that: 1) \eqref{eq:pde} 
		is a linear system of
		equations and 2) the homogeneity of \eqref{eq:pde-diffusion} implies that the movement of free particles is 
		independent in the three spatial dimensions.
		It follows then that the location of a particle originally released at the origin $\pos = \mathbf{0}$, will, 
		after a time $\tau~[\mathrm{s}]$
		in Brownian motion with no intermediate binding events, have a distribution over the $xy$-plane given by the 
		Green 
		function for the homogeneous diffusion equation in 2D during a time $\tau$, i.e., $g_\sigma$ as in 
		Definition~\ref{def:GaussianKernel} with 
		$\sigma = \sqrt{2D\tau}~[\mathrm{m}]$ (see \cite{Ursell2007}). 
		\begin{definition}[Gaussian kernels] \label{def:GaussianKernel}
			$\lbrace g_\sigma: \reals^2\rightarrow \reals_+ \rbrace_{\sigma>0}$ is a scale family of 2D rotationally 
			invariant Gaussian kernels, where
			\begin{IEEEeqnarray*}{c} \label{eq:Gaussian}
				g_\sigma(\pos) = \frac{1}{2\pi \sigma^2}\exp\left(-\frac{\pos^{\mathrm{T}} \pos}{2\sigma^2}\right), 
				\,\forall \pos \in \reals^2\,.
			\end{IEEEeqnarray*}
		\end{definition}

		Because Brownian motions are Markov processes, it follows that the total displacement over directions in the 
		$xy$-plane is fully determined by the total time
		in free motion $\tau$, even when intermediate binding events are present.
		The total time in free motion of any given particle over some specific time interval is, however, random. 
		Specifically, it depends on the particle's random trajectory, on 
		the collisions of the trajectory with the plane $z=0$, and on the subsequent random associations (adsorption) and 
		disassociations (desoprtion), as modeled by \eqref{eq:pde-surface} and
		\eqref{eq:pde-boundaryflow}. Let 
		\begin{IEEEeqnarray}{c} \label{eq:varphi}
			\varphi:\lbrace (\tau,t)\in[0,T]^2 \mid \tau\leq t\rbrace \rightarrow \reals_+\,,
		\end{IEEEeqnarray}
		such that $\varphi(\tau,t)$ is the 
		probability (density)\footnote{The terms \emph{probability} or \emph{probability density} are
		technically incorrect in this case. A formal definition of the quantity $\varphi(\tau,t)$ is given in 
		Section~\ref{ssec:data-synthesis}.} 
		of a particle being in free motion for a total time $\tau$ before being found in a bound state at time 
		$t~[\mathrm{s}]$.
		$\varphi(\tau,t)$ is defined for $\tau \in [0,t]$ and is determined implicitly by \eqref{eq:pde}.
		The specific nature of $\varphi(\tau,t)$ is of little relevance to the objective of this section, and only its 
		existence is required. Nonetheless, its characterization will be fundamental for some of the uses of our
		observation model.
		In Section~\ref{ssec:data-synthesis}, we present a novel and detailed derivation of 
		$\varphi$ in terms of $\ka$, $\kd$ and $D$, extending results from \cite{Berezhkovskii2004} by 
		characterizing desorption from the surface in terms of its effect on the total time in free motion.
		
		For a given $\varphi$, then, the spatial probability density of finding a particle bound to the surface at 
		time $t$, after a release into the medium at the origin 
		at time $0$, is given by (see \cite{Berezhkovskii2004})
		\begin{IEEEeqnarray*}{c}
			p(\pos,t) = \int_0^t g_{\sqrt{2D\tau}}(\pos) \varphi(\tau,t) \dint\tau \, ,
		\end{IEEEeqnarray*}
		which can be viewed as the Green function for $d(\pos,t)$ in \eqref{eq:pde}.
		Note that $p(\pos,t)$ integrated over $\pos\in\reals^2$ yields the probability that a particle released at 
		time $0$ is bound at time $t$.
		By linearity, time-invariance, and spatial invariance on the directions in the $xy$-plane, it follows that 
		$d$ can be expressed as
		\begin{IEEEeqnarray}{c} \label{eq:density}
			d(\pos,t) = (s*p)(\pos,t)\,,
		\end{IEEEeqnarray}
		i.e., as a spatio-temporal convolution of the Green function $p$ and the SDR $s$.
		For the spatial part of the convolution in \eqref{eq:density}, it is convenient to introduce the Gaussian blur
		operators as follows.
		\begin{definition}[Gaussian blur operators] \label{def:GaussBlur}
			\begin{IEEEeqnarray*}{c}
				\left\lbrace G_\sigma\in\opers{\LebTwo{\reals^2}}{\LebTwo{\reals^2}} \right\rbrace_{\sigma >0}
			\end{IEEEeqnarray*}
			is a family of convolutional operators, where
			\begin{IEEEeqnarray*}{c}
				(G_\sigma f)(\pos) = \int_{\reals^2} f(\pos-\boldsymbol{\rho}) g_\sigma(\boldsymbol{\rho}) \dint\boldsymbol{\rho}, \forall f\in \LebTwo{\reals^2},
			\end{IEEEeqnarray*}
			and $g_\sigma$ is given by Definition~\ref{def:GaussianKernel}.
		\end{definition}

		By using Definition~\ref{def:GaussBlur} and evaluating the convolution in \eqref{eq:density} independently over
		the spatial and temporal dimensions,
		we can express $d$ compactly as 
		\begin{IEEEeqnarray}{c} \label{eq:density2}
			d(\pos,t) = \int_0^t G_{\sqrt{2D\tau}} \, v(\pos, \tau,t) \dint\tau \, ,
		\end{IEEEeqnarray}
		where $v:\reals^2\times \reals_+^2 \rightarrow \reals_+$ is such that
		\begin{IEEEeqnarray}{c} \label{eq:weights-w}
			v(\pos,\tau,t) = \int_{\tau}^t s(\pos,t-\eta) \varphi(\tau,\eta) \dint\eta \,.
		\end{IEEEeqnarray}
		$v$ summarizes the effect of movement in the $z$-dimension, adsorption, and desorption on the diffusion of the
		particles generated with a
		source density rate $s$. Theorem~\ref{theorem:2D-DataModel}	summarizes the conclusions from the discussion above
		in terms of the image 
		observation $\obs$. A step-by-step derivation of \eqref{eq:density2}, 
		\eqref{eq:weights-w}, \eqref{eq:observation-model}, and \eqref{eq:weights-a} from
		\eqref{eq:density} can be found in the supplementary material to this paper.
			
		\begin{theorem}[Observation model] \label{theorem:2D-DataModel}
			Let $\obs:\reals^2\rightarrow \reals_+$ be the spatial density of bound particles at time $T$, i.e., when 
			the experiment finishes. Then, we have that
			\begin{IEEEeqnarray}{c} \label{eq:observation-model}
				\obs(\pos) = \int_0^{\sigmax} G_{\sigma} \, a(\pos,\sigma) \dint\sigma \, ,
			\end{IEEEeqnarray}
			where $\sigma = \sqrt{2 D \tau}$, $\sigmax = \sqrt{2DT}$, and
			\begin{IEEEeqnarray}{c} \label{eq:weights-a}
				a(\pos,\sigma) = \frac{\sigma}{D} v\!\left( \pos , \frac{\sigma^2}{2D}, T \right) \,,
			\end{IEEEeqnarray}
			with $v$ as in \eqref{eq:weights-w}.
			We will refer to $a:\Omega\rightarrow \reals_+$ in \eqref{eq:weights-a} as the post adsorption-desorption 
			source density rate (PSDR).
		\end{theorem}
		
		An important feature of \eqref{eq:observation-model} is that the spatial properties of $s$ are retained by $a$,
		as \eqref{eq:weights-w} and 
		\eqref{eq:weights-a} operate only on the temporal dimension. 
		This implies that the PSDR $a$ contains the same amount of information 
		for SL as the original SDR $s$.
		
		The value of the PSDR $a(\pos,\sigma)$ can be interpreted as the density of particles released from a location 
		$\pos$
		that will appear in $\obs$ after a 2D diffusion of $\tau=\sigma^2/(2D)$. 
		The model in \eqref{eq:density2}, however, is not a 2D diffusion model with an equivalent source $v$, due to the 
		dependence of $v$ on the 
		observation time $t$. Nonetheless, we can and will treat \eqref{eq:observation-model} as our observation model
		with $a$, rather than $s$,
		as the sought unknown quantity. This will result in a few important benefits. First, the relative 
		simplicity of \eqref{eq:observation-model} will 
		prove beneficial both for formulating the inverse diffusion problem and for constructing its algorithmic
		solution. Second, the recovery of
		$a$ in \eqref{eq:weights-a} can be addressed without an explicit characterization of $\varphi$. This removes 
		the need for the values of
		$\ka$, $\kd$ and $D$, which can vary between assays, are costly to measure \cite{Matthes2005}, and hard to
		estimate \cite{Schwab2012}.
		This said, an explicit characterization of $\varphi$ is still desirable. In particular, it 
		could enable the recovery of the 
		original SDR $s$ from the recovered $a$, and it allows for simulation of data based on a given SDR $s$, which 
		is easier to postulate than the PSDR $a$. We therefore continue by providing an explicit characterization of 
		$\varphi$ that exhibits favorable properties with regards to its numerical approximation.
		
	\subsection{Physical parameters and data synthesis} \label{ssec:data-synthesis}
		
		The quantity $\varphi(\tau,t)$ summarizes the relation between the time $t$ at which a particle released at 
		time $0$ is found bound, and the total
		time $\tau$ it has spent in free movement. Formally, consider a particle released at time $0$ and let its 
		position in the $z$-dimension be 
		$\lbrace z_t \rbrace_{t\in[0,T]}$.  For each $t\in[0,T]$, consider the random variables 
		$\tau= |\lbrace \tilde{\tau} \in [0,t] : z_{\tilde{\tau}} > 0\rbrace|$, i.e., 
		the time in free motion before $t$, and $b_t\in\lbrace 0, 1\rbrace$ such that $b_t=1$ if $z_t=0$ and $b_t=0$ 
		otherwise, i.e., an indicator of the 
		particle being bound\footnote{Note here that the event $z_t=0$ is equivalent to the particle being bound 
		($b_t=1$) because under a free Brownian motion,
		$z_t=0$ has probability $0$.}
		at time $t$. Then, $\varphi(\tau,t)$ is formally
		the Radon-Nikodym derivative of the joint distribution of the continuous random variable $\tau$ and the 
		discrete random variable $b_t$, 
		i.e., $\forall \tau \in [0,t]$,
		\begin{IEEEeqnarray*}{c}
			\varphi(\tau,t) = f_{\tau|b_t}(\tau|b_t=1) \Pr\left(b_t=1\right)\,,
		\end{IEEEeqnarray*}
		where $f_{\tau|b_t}(\cdot|b_t=1)$ is the probability density function of the time in free motion $\tau$ given that 
		the particle is found bound at time $t$.
		
		Obtaining a characterization of $\varphi(\tau,t)$ in terms of $\ka$, $\kd$ and $D$ provides further possibilities
		to exploit the model \eqref{eq:observation-model}.
		First, one can use this model to obtain synthetic data that corresponds to specific 
		reaction-diffusion-adsorption-desorption models and specific 
		source density rates $s(\pos,t)$, which provides a way of objectively comparing algorithmic proposals. Second, 
		one could, if the physical parameters of a real assay were known,
		address the inverse problem of obtaining $s(\pos,t)$ from any estimation of $a(\pos,\sigma)$ by inverting the 
		linear system formed by equations \eqref{eq:weights-w} and
		\eqref{eq:weights-a}. In Theorem~\ref{th:charvarphi-general}, we provide the full 
		characterization of 
		$\varphi(\tau,t)$ in terms of the physical parameters of the model. Both Theorem~\ref{th:charvarphi-general} and
		Lemma~\ref{lem:charvarphi-noescape}, an intermediate
		result, are proved in Appendix~\ref{app:PhyPar}. In Fig.~\ref{fig:exampleSIMdata}, we show a section of a synthetic image generated 
		using the result in Theorem~\ref{th:charvarphi-general}, for comparison with the image obtained from a real Fluorospot 
		assay in Fig.~\ref{fig:physics}.
		\begin{figure}
			\centering
			\includegraphics[keepaspectratio=true,scale=1]{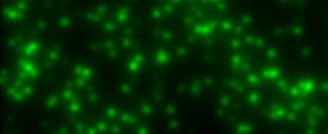} \\
			\caption{ 
					Example section of an image simulated from model \eqref{eq:pde} using the result in 
					Theorem~\ref{th:charvarphi-general}, observed through a simulated imperfect, noisy image acquisition
					system. 
					The image is loaded into the green channel for ease of comparison with Fig.~\ref{fig:physics}.
					For details on the discretization and numerical techniques employed to obtain this image, see 
					Section~\ref{sec:Disc}, Section~\ref{ssec:data-synthesis}, and the supplementary material to this paper.
					\label{fig:exampleSIMdata}
					} 
		\end{figure}
		
		Consider first a characterization of the simpler case $\kd=0$, i.e., the case in which particles that are 
		bound to the 
		surface can not be desorbed, which we present in Lemma~\ref{lem:charvarphi-noescape}.
		\begin{lemma}[Characterization of the observation model from physical parameters, Case $\kd=0$] 
		\label{lem:charvarphi-noescape}
			Consider model \eqref{eq:pde} when $\kd=0$. Then, we have that
				$\varphi(\tau,t) =  i_{[0,t)}(\tau) \phi(\tau)$,
			with
			\begin{IEEEeqnarray}{rl}\label{eq:first-adsorption}
				\phi(\tau) =  \frac{\ka}{\sqrt{\pi D \tau}} 
				- \frac{\ka^2}{D} \mathrm{erfcx}\left( \ka \sqrt{\frac{\tau}{D}} \right) \,, 
			\end{IEEEeqnarray}
			where
			\begin{IEEEeqnarray*}{c}
				\mathrm{erfcx}(x) = e^{x^2} \mathrm{erfc}(x) \,\mbox{, and }
				\mathrm{erfc}(x)= \frac{2}{\sqrt{\pi}} \int_x^\infty e^{-t^2} \dint t\,,
			\end{IEEEeqnarray*}
			are the scaled-complementary and complementary error functions, respectively.
		\end{lemma}
		Theorem~\ref{th:charvarphi-general} extends this result to the general case in which $\kd\geq0$ by segmenting
		the total time of free motion in 
		subsequent fractions of free motion interrupted by adsorption-desorption events. 
		\begin{theorem}[Characterization of the observation model from physical parameters] \label{th:charvarphi-general}
			Consider model \eqref{eq:pde}. Then, we have that
			\begin{IEEEeqnarray}{c} \label{eq:sum-varphi}
				\varphi(\tau,t) = i_{[0,t)}(\tau) \sum_{j=1}^\infty \phi^{j*}(\tau) p\left[j-1;\kd(t-\tau)\right] \, ,
			\end{IEEEeqnarray}
			where $\phi^{j*}(\tau)$ is the $j$-th convolutional power of $\phi(\tau)$ in \eqref{eq:first-adsorption} and
			\begin{equation}\label{eq:Poisson}
				p[j;\lambda] = \frac{\lambda^j e^{-\lambda}}{j!}, \forall j \in \natu,\forall \lambda \geq 0\,,
			\end{equation}
			is the probability mass function of a Poisson random variable with mean $\lambda$ evaluated at $j$.
		\end{theorem}
			
		Note that \cite{Agmon1984} also studied model \eqref{eq:pde} with $\kd\geq0$, but in terms of an expression 
		for the distribution of particles
		in the $z$-dimension after a time $t$ since they were released, i.e. a parallel to the $u(z,t)$ used in  
		Appendix~\ref{app:PhyPar} to prove Lemma~\ref{lem:charvarphi-noescape}.
		However, because our goal is to characterize the total time in free motion, and $u(z,t)$ does not reveal how a 
		particle arrived
		at a position $z$ at time $t$, our result in Theorem~\ref{th:charvarphi-general} is needed.
		
		For any practical application, $\varphi(\tau,t)$ needs to be computed, i.e. approximated and discretized in some 
		manner. 
		To this end, note that truncating \eqref{eq:sum-varphi} at a finite $J_\epsilon$ can be done at any arbitrary error level
		$\epsilon>0$.
		Intuitively, this is because $p[j;\lambda]$ decays exponentially for large $j$s while $\phi^{j*}(\tau)$ is 
		stable (in norm) with $j$. 
		Formally, this result is stated in Lemma~\ref{lem:Truncate} and proved in Appendix~\ref{app:PhyPar}.
		\begin{lemma}[Truncation of the sum to characterize the model] \label{lem:Truncate} 
			Consider, for any $\epsilon > 0$, 
			\begin{IEEEeqnarray*}{c}
				J_\epsilon = Q_{\mathrm{Poi}}\left(1-\frac{\epsilon}{\normLebTwo{0,+\infty}{\phi}^2};\kd T\right)\,,
			\end{IEEEeqnarray*}
			where $Q_\mathrm{Poi}(p;\lambda)$ is the quantile function, i.e. the inverse cumulative distribution function,
			of a Poisson random variable with mean $\lambda>0$ evaluated at $p\in(0,1)$. Then, 
			\begin{IEEEeqnarray*}{c}
				\tilde{\varphi}(\tau,t) = 
				\left| \varphi(\tau,t) - \sum_{j=1}^{J_\epsilon-1} \phi^{j*}(\tau) p\left[j-1;\kd(t-\tau)\right] \right|
				\leq \epsilon\,,
			\end{IEEEeqnarray*}
			$\forall  (\tau,t) \in [0,T]^2$ such that $\tau\leq t$.
		\end{lemma}	
		
		Finally, note that a discrete approximation to the $j$-th convolutional power $\phi^{j*}(\tau)$ can be computed
		numerically by discretization of $\phi(\tau)$ and recursive discrete convolution.
	
	\subsection{Model flexibility and Imaging limitations} \label{ssec:SensLim}
	
		To assume that an imaging system can provide measurements given by \eqref{eq:observation-model} is, naturally, an
		idealization. Besides finite dimensionality, which will be treated in detail in Section~\ref{sec:Disc}, any 
		physically feasible image acquisition must deviate from this model due to the following factors: 1) random 
		photonic and electronic events, often modeled by additive white noise in the final observation, 2) non-linear 
		effects, such as limited dynamic ranges or quantization, 3) linear effects, such as a blur with a point-spread
		function (PSF) that limits the resolution of a particular optical system, and 4) a bounded field of view.
		
		In terms of the effect of random events, we will numerically demonstrate in Part II of this 
		paper~\cite[Section III]{AguilaPla2017a} that our approach to SL is very robust to the presence of additive white 
		noise.
		In terms of non-linear effects, we assume that the dynamic range of the camera is adjusted automatically so that
		saturation is not an issue. Additionally, we implicitly assume that the imaging sensor is noise-limited instead 
		of quantization-limited, as is the case with most current cameras. In our empirical validation on synthetic data in 
		Part II~\cite[Section III]{AguilaPla2017a}, we enforce this by using a statistical model for 
		quantization to ensure that the levels of additive noise under analysis are much larger than those expected from 
		quantization error in a current scientific camera, e.g., $12$-bit quantization.
		In terms of linear effects, we will assume that the PSF for the optical system is monomodal and symmetric, and well approximated by
		a Gaussian kernel $g_{\sigma_\mathrm{b}}$ with some standard deviation $\sigma_{\mathrm{b}}$. 
		Under these assumptions, we modify \eqref{eq:observation-model} to express the blurred 
		observation $\obs^{\mathrm{b}}$ as
		\begin{IEEEeqnarray}{rl} \label{eq:blur-observation-model}
			\obs^{\mathrm{b}}=g_{\sigma_\mathrm{b}} * \obs & = \int_0^{\sigmax} G_{\sigma+\sigma_\mathrm{b}} a_\sigma \dint\sigma \,.
		\end{IEEEeqnarray}
		All of the results in this two-part paper are invariant to this shift in $\sigma$ and can 
		be re-derived mutatis mutandis for \eqref{eq:blur-observation-model}. Finally, in terms of the limited field of view, we will make the reasonable assumption that 
		all the sources we aim at recovering are within the camera's field of view.
		
	\ifbool{tot}{}{
		\bibliographystyle{IEEEtran}
		\bibliography{IEEEabrv,\bib/multi_deconv}
	}

		\section{Inverse Diffusion by Functional Optimization} \label{sec:InvDif}

	\subsection{Optimization problems for inverse diffusion} \label{sec:InvDif:Opti}
	
		In this section, we first present the inverse problem of recovering the PSDR $a$
		from the density of bound analyte $\obs$ as a non-negative minimum-norm functional optimization problem.
		Then, we propose to address the ill-posedness of this naïve minimum-norm formulation by 
		regularizing it according to the available prior knowledge. As a result, we propose a non-negative group-sparsity regularized minimum-norm optimization 
		problem to fit the observation model in Section~\ref{sec:DataModel} to the data.
		
		Our treatment and language will be that of functional analysis, which will enable the exposition of 
		the optimization problem in the natural spaces of particle densities.
		Although discretization will eventually be necessary for the synthesis and analysis of data, introducing it already in the observation model 
		\eqref{eq:observation-model} would mask the generality of the proposed approach. Indeed, in Section~\ref{sec:Disc} we propose a simple
		discretization scheme for \eqref{eq:observation-model} and any functional algorithm for inverse diffusion, but our exposition opens up inverse diffusion to the 
		use of more sophisticated discretization schemes. For example, off-the-grid solutions such as \cite{Catala2017} dynamically estimate 
		the support of a discrete measure observed through a convolutional operator, and offer opportunities for more rigorous mathematical analysis.
		In conclusion, in a philosophy strongly supported by \cite[ch. 5]{Tarantola2005}, we present an optimization problem to address inverse diffusion on a 
		functional (infinite-dimensional) setting and discretize the problem only after that. Similarly, in Part~II~\cite{AguilaPla2017a}, we propose first the 
		functional version of the algorithm, and use the simple discretization in Section~\ref{sec:Disc} only after that to provide an implementable algorithm and empirical results.
				
		We begin by introducing the Hilbert spaces needed to properly state the inverse problem. The definitions of these function spaces will permit 
		the adaptation of the problem to specific, practical conditions. For instance, in defining
		the space of observed densities, we include a weighting function that enables us to, in each case, set a value on the cost of wrongly predicting 
		the observed density in each location. This allows us, for example, to make the inverse problem robust to regions of an 
		image sensor that are known beforehand to be faulty or irrelevant.
		\begin{definition}[Observed density space] \label{def:DenSpace}
			Consider a weighting function $w\in \mathrm{L}_+^\infty\left(\reals^2\right)$ such
			that $w \neq 0$. Then, the bilinear form 
			\begin{IEEEeqnarray*}{c}
				\prodDen{d_1}{d_2} = \int_{\reals^2} w^2(\pos) d_1(\pos) d_2(\pos) d\pos,\, \forall d_1,d_2:\reals^2 \rightarrow \reals\,,
			\end{IEEEeqnarray*}
			is positive and symmetric.
			Therefore, the linear space 
			\begin{IEEEeqnarray*}{c}
				\DenSpace=\left\lbrace d:\reals^2\rightarrow \reals : \prodDen{d}{d} < +\infty \right\rbrace\,,
			\end{IEEEeqnarray*}
			equipped with the inner product $\prodDen{\cdot}{\cdot}$ is a Hilbert space, and it is where the observed density lies, 
			i.e., $\obs \in \DenSpace_+$.
		\end{definition}
		Similarly, in defining the space of PSDRs, we include a masking pattern that indicates which locations can hold cells and which cannot. This 
		reduces the support of the considered PSDRs, thus making the inverse problem easier by incorporating prior knowledge.
		\begin{definition}[PSDR space] \label{def:ASpace}
			Consider a masking pattern function $\mu:\reals^2\rightarrow \left\lbrace 0,1\right\rbrace$ with a non-empty bounded support
			$\supp{\mu}$.
			Then, the linear space
			\begin{IEEEeqnarray*}{c}
				\ASpace = \left\lbrace 
				a \in  \LebTwo{\Omega} : \supp{a} \subseteq \supp{\mu} \times [0,\sigmax] 
				\right\rbrace\,,
			\end{IEEEeqnarray*}
			equipped with the inner product $\prodA{\cdot}{\cdot} = \prodLebTwo{\Omega}{\cdot}{\cdot}$ is a Hilbert space, and
			it is the space where the PSDR lies, i.e., $a\in\ASpace_+$. Here, recall that $\Omega = \reals^2 \times \reals_+$.
		\end{definition}
		
		The core of the observation model in Theorem~\ref{theorem:2D-DataModel} is the operator that reflects how a change in the PSDR $a\in\ASpace_+$ affects the
		observed density  $\obs \in \DenSpace_+$, i.e., the observation operator in this inverse problem. We refer to it as the diffusion operator
		because of the parallelism between $\obs$ and an observation of a 2D diffusion process with SDR $v$, which we discussed at the end of Section 
		\ref{sec:DataModel:ObsModel}.
		\begin{definition}[Diffusion operator] \label{def:DiffOp}
			The linear operator $A: \ASpace \rightarrow \DenSpace$ such that
			\begin{IEEEeqnarray*}{c}
				A a = \int_{0}^{\sigmax} G_\sigma a_\sigma \dint\sigma,\, \forall a \in \ASpace\,,
			\end{IEEEeqnarray*}
			represents the dependence between $a$ and $\obs$ specified by Theorem~\ref{theorem:2D-DataModel}. Here,
			$a_\sigma : \reals^2 \rightarrow \reals_+$ is such that $a_\sigma(\pos) = a(\pos,\sigma)$, $\forall (\pos,\sigma)\in\Omega$.
		\end{definition}
			
		The measurement model \eqref{eq:observation-model} can now be succinctly expressed as $\obs = A a$.
		In this view, the estimation of the PSDR $a$ may be addressed as a least squares 
		problem with respect to the operator $A$, i.e., as the convex optimization problem
		\begin{IEEEeqnarray}{c} \label{eq:InvDif:NonRegularised:IndicatorNotation}
			\min_{a\in\ASpace} \left[ \normDen{Aa - \obs}^2 + \delta_{\ASpace_+}(a) \right]\,.
		\end{IEEEeqnarray}
		Here, the penalty function for the prediction $Aa$ is precisely the square of the norm $\normDen{\cdot}$ on the space of observations, which takes
		into account the weighting $w$ to determine the importance of an error in each of the different spatial positions. 
		Additionally, for the PSDR to have physical meaning, the positivity constraint $a\in \ASpace_+$ has to be met.
		Note that $\ASpace_+ \subset \ASpace$ is a convex cone, and that the indicator function notation for the convex constraint is
		convenient for later treatment.
		
		Because the dimensionality of the PSDR $a$ exceeds that of the observation $\obs$, \eqref{eq:InvDif:NonRegularised:IndicatorNotation} is ill-posed in the 
		sense that many different PSDRs lead to the same observation. This calls for the use of a regularizer that eases the inverse problem by 
		biasing the solution towards more plausible explanations. In particular, we propose to use a non-negative group-sparsity regularizer to induce
		group behavior in the $\sigma$-dimension and sparsity in the spatial dimensions. 
		This is consistent with the explanation of the bound density $\obs$ as a result of particle generation by a finite number of spatially separated,
		immobilized cells. Therefore, we propose to solve the optimization problem
		\begin{IEEEeqnarray}{c} \label{eq:InvDif:Regularised}
			\!\!\!\!\!\!\!\min_{a\in\ASpace} \left[ \normDen{Aa - \obs}^2 \!+\! \delta_{\ASpace_+}(a) \!+\! 
			\lambda \normLebOne{\reals^2}{\normLebTwo{\reals_+}{\xi a_\pos}} \right]
		\end{IEEEeqnarray}
		which is convex and suited to iterative non-smooth convex optimization methods, as we show in Part~II~\cite{AguilaPla2017a}. Here, for each $\pos\in\reals^2$,
		${a_\pos : [0,\sigmax] \rightarrow \reals_+}$ is such that $a_\pos(\sigma) = a(\pos,\sigma)$ for any $\sigma\in[0,\sigmax]$, $\lambda>0$ is the
		regularization parameter, and $\xi\in \mathrm{L}_+^\infty[0,\sigmax]$ is a non-negative bounded weighting function in $\sigma$ that can
		be used to incorporate further prior knowledge. For example, if one knew
		the exact parameters $\ka$, $\kd$ and $D$ of the physical system, one could use the characterization of 
		$\varphi$ in Theorem~\ref{th:charvarphi-general} 
		to choose $\xi$ so that the penalization in \eqref{eq:InvDif:Regularised}
		corresponds to a uniform penalization through $t$ in the original SDR $s$. 
		Additionally, if one knew that the particular experimental setting only allows for cells to generate 
		particles during times $t$ such that $t_0<t<t_1<T$, one could choose $\xi$ to have very large values for
		$\sigma\in[0,\sqrt{2D t_1}] \cup [\sqrt{2D(T-t_0)}, \sigmax]$. Finally, if one wanted to impose the restrictions
		of the model only on a certain range of $\sigma$s, say $\sigma\in\aleph\subset[0,\sigmax]$, and relax them for its complement 
		$\complement{\aleph}=[0,\sigmax]\setminus \aleph$, one could choose $\xi$ such that $\supp{\xi}=\aleph$.
		The case in which $\xi$ is simply the $(0,1)$-indicator function of a set $\aleph$ is of special relevance in Part~II~\cite{AguilaPla2017a} due to its
		tractability, and is useful, for example, to use the values of $a$ for 
		$\sigma \in \complement{\aleph}$
		to account for a low-frequency background that could not be explained by cell secretion alone. 
		
		Note now that while in both \eqref{eq:InvDif:NonRegularised:IndicatorNotation} and \eqref{eq:InvDif:Regularised} we have used $\min$ instead of $\inf$, we have yet
		been unable to formally prove that these problems do have a minimizer in $\ASpace_+$. Nonetheless, in the following section we provide some results that characterize the diffusion operator 
		$A$, providing some insight on its structure and beneficial properties. Some of these results will enable us to prove, in Section~\ref{sec:Disc}, that the discretized 
		equivalents to \eqref{eq:InvDif:NonRegularised:IndicatorNotation} and \eqref{eq:InvDif:Regularised} under our discretization scheme do have a minimizer. 
		
	\subsection{Characterization of the diffusion operator}
	
			First, we verify that $A$ is a continuous, i.e., bounded, linear operator. 
			Although this does not provide the existence of a minimizer of \eqref{eq:InvDif:NonRegularised:IndicatorNotation} or \eqref{eq:InvDif:Regularised},
			it does give some intuitive hope in terms of the bounded inverse theorem.
			\begin{lemma}[Boundedness of the diffusion operator] \label{lem:NormDiff}
				The norm  in $\opers{\ASpace}{\DenSpace}$ of the linear operator $ A:\ASpace\rightarrow \DenSpace$ in Definition~\ref{def:DiffOp}
				is bounded as
				\begin{IEEEeqnarray*}{c}
					\|A\|_{\opers{\ASpace}{\DenSpace}} \leq \sqrt{\sigmax} \wninf\,.
				\end{IEEEeqnarray*}
				Thus, $A$ is a bounded operator and, because $A$ is linear, $A$ is a linear continuous operator, i.e.
				$A\in\opers{\ASpace}{\DenSpace}$. 
			\end{lemma}
			
			We proceed by characterizing the nullspace of the operator, showing that it only contains a very specific class of functions.
			This rather simple result, which is also valid for any convolutional operator with non-negative unit $\mathrm{L}^1$-norm kernel, will be of great help
			when characterizing the existence of minimizers in the discrete case.
			\begin{lemma}[Nullspace of the diffusion operator] \label{lem:NullSpace}
			  Consider the nullspace of the diffusion operator, i.e., $\NullSpace{A}=\lbrace a \in \ASpace : Aa = 0\rbrace$. 
			  Then, $\forall a \in \NullSpace{A}$, we have that $\normLebOne{\Omega}{a_+} = \normLebOne{\Omega}{a_-}$.
			\end{lemma}
			Note that an immediate consequence of Lemma~\ref{lem:NullSpace} is that $\ASpace_+\cap \NullSpace{A}=\lbrace0\rbrace$, since 
			$a\in\ASpace_+$ implies $\normLebOne{\Omega}{a_-}=0$, which, if $a\in\NullSpace{A}$, implies that $\normLebOne{\Omega}{a_+}=0$, i.e., $a=a_++a_-=0$.
			
			For the sake of completeness, we also present here the expression for the adjoint operator $A^*$ of the diffusion operator $A$,
			which will be of great value in the design of algorithms to minimize \eqref{eq:InvDif:Regularised} in 
			Part~II~\cite{AguilaPla2017a}.
			\begin{lemma}[Adjoint to the diffusion operator] \label{lem:AdjDiff}
				The adjoint operator $A^*\in\opers{\DenSpace}{\ASpace}$ to the diffusion operator $A$ in Definition~\ref{def:DiffOp} is 
				such that
				\begin{IEEEeqnarray*}{c}
					\left( A^*d \right)(\pos,\sigma) = \mu(\pos) \cdot \left(G_\sigma \left\lbrace w^2 d \right\rbrace \right)(\pos), \forall d \in \DenSpace\,,
				\end{IEEEeqnarray*}
				for each $\pos\in\reals^2$ and $\sigma>0$. 
			\end{lemma}
			
			Both Lemma~\ref{lem:NormDiff} and Lemma~\ref{lem:AdjDiff} are based on an equivalent 
			characterization of the family of Gaussian blur operators $\left\lbrace G_\sigma \right\rbrace$ from Definition~\ref{def:GaussBlur}, based 
			on standard results for convolutional operators. This characterization, i.e., Properties \ref{prop:NormOneGauss} and \ref{prop:SelfGaus},
			can be found together with the proof for the results in this section in Appendix~\ref{app:DiffOp}.
			
	\ifbool{tot}{}{
		\bibliographystyle{IEEEtran}
		\bibliography{IEEEabrv,\bib/multi_deconv}
	}

		\section{Discretization} \label{sec:Disc}

			In practice, no imaging sensor is infinitely resolute. A digital camera will
			instead obtain a matrix $\dobs \in \matricesP{M,N}$, which will be some discretization of $\obs$ in \eqref{eq:observation-model}. 
			Here, $M,N\in\natu$ represent
			the number of pixels in each dimension of the imaging sensor, with typical values of $1024$ or $2048$ for current 
			scientific cameras.  
			In particular, the relation between $\dobs$ and $\obs$ in a generic discrete imaging sensor can 
			be modeled by the operator $\Restr{\DenSpace}: \DenSpace \rightarrow \matrices{M,N}$ such that
			\begin{IEEEeqnarray}{c} \label{eq:Rd}
				\tilde{d} = \Restr{\DenSpace}\left(d\right) = \left\lgroup \tilde{d}_{m,n} \right\rgroup = 
				\left\lgroup \int_{\discreg{m,n}} d(\pos)\,\dint\pos \right\rgroup, 
			\end{IEEEeqnarray}
			$\forall d \in \DenSpace$, where $m\in\listint{M}$, $n\in\listint{N}$ and
				$\discreg{m,n} = \left\lbrace (m,n) \right\rbrace + [-0.5,0.5]^2$ 
			is the region that corresponds to the pixel at the position $(m,n)$. Here, the scale of each spatial variable is 
			normalized with respect to the pixel size to simplify further derivations. Additionally, each spatial variable is 
			translated so that $(1,1)$ corresponds to the location of the first pixel's center. Note that in order to preserve consistency,
			normalization is also needed on the $\sigma$-dimension in \eqref{eq:observation-model}--\eqref{eq:weights-a}, i.e., we will work with 
			$\dsigma = \sigma / \Delta_{\mathrm{pix}}$ and $\dsigmax = \sigmax / \Delta_{\mathrm{pix}}$,
			where $\Delta_{\mathrm{pix}}~[\mathrm{m}]$ is the length of a pixel's side.
		
			In the classical theory of discretization~\cite[Ch.~34--35]{Zeidler1990}, one wants to numerically solve a functional inverse 
			problem, e.g., find $a\in\ASpace_+$ such that $A a = \obs$ for a specific $\obs\in\DenSpace_+$. Then, to do so numerically,
			one defines a discretization scheme parametrized by the dimensionalities $q_1,q_2\in\natu$ of the observation
			and solution, i.e., some rule to obtain approximations 
			\begin{IEEEeqnarray*}{c} 
				\widetilde{A}_{q_1,q_2} \in \matrices{q_1,q_2},\,
				\tilde{a}_{q_1} \in \reals^{q_1},\, \dobsq \in \reals^{q_2},
			\end{IEEEeqnarray*}
			and one solves $\widetilde{A}_{q_1,q_2} \tilde{a}_{q_1} = \dobsq$ instead, relying on equivalence results when 
			$q_1,q_2 \rightarrow +\infty$. 
			In our case, we want to solve the functional optimization problem proposed in~\eqref{def:DiffOp},
			but only have access to a discretized observation $\dobs = \Restr{\DenSpace}\left( \obs \right)$. This imposes the structure in \eqref{eq:Rd}
			onto our discretization of the image observation $\obs$, and fixes its dimension to $q_2 = M\times N$.
			With respect to \eqref{eq:InvDif:Regularised}, we will assume that the user-specified parameters $\mu$, $w$ and $\xi$ are 
			chosen consistently with the discretization. For example, we will assume that instead of a weighting function $w(\pos)$, we have a weighting matrix
			$\tilde{w}=\Restr{\DenSpace}(w)\in\matrices{M,N}$, at the same discretization level as $\dobs$.
			
		Classical results in discretization theory~\cite{Zeidler1990} cover only cases in which the functional inverse problem is well-posed.
		In fact, the design of discretization schemes in the context of possibly ill-posed inverse problems, such as 
		\eqref{eq:InvDif:Regularised}, is an open research topic~\cite{Mathe2003,Kaipio2007,Erb2015,Haemarik2016}. 
		Thus, formulating a discretization that is optimal in some sense is beyond 
		the scope of this paper. Instead, we will use the basic ideas from inner approximation schemes~\cite[Ch.~34]{Zeidler1990} to 
		propose an intuitively natural discretization. 
		
		A discretization scheme for \eqref{eq:InvDif:Regularised} under the inner approximation paradigm involves two restriction operators, i.e. 
		\begin{IEEEeqnarray*}{c}
			\Restr{\ASpace}: \ASpace \rightarrow \ASpace_{q_1}\,,\,\,
			\Restr{\DenSpace}: \DenSpace \rightarrow \DenSpace_{q_2}\,,
		\end{IEEEeqnarray*}
		where $\ASpace_{q_1}$ and $\DenSpace_{q_2}$ are $q_1$- and $q_2$-dimensional spaces, respectively, and $\Restr{\DenSpace}$ is 
		characterized in \eqref{eq:Rd} with $\DenSpace_{q_2} = \matrices{M,N}$ and two 
		extension operators, i.e. 
		\begin{IEEEeqnarray*}{c}
			\Exten{\ASpace}:  \ASpace_{q_1}  \rightarrow \ASpace\,,\,\,
			\Exten{\DenSpace}: \DenSpace_{q_2} \rightarrow \DenSpace \,.
		\end{IEEEeqnarray*}
		These operators fully characterize the discretization scheme, because they not only determine the discrete approximation
		of each element in $\ASpace$ or $\DenSpace$ through the restriction operators, but also the discrete approximation of any 
		operator from and to these spaces.
		In our case, we are specifically interested in the finite-dimensional approximation of the diffusion operator $A$ under a given discretization 
		scheme, which can be used directly to synthesize data, but also will be a fundamental step in any discrete iterative procedure to approximate a 
		solution to \eqref{eq:InvDif:Regularised}. The latter also applies to finding the discrete expression for the adjoint $A^*$, which
		will play an important role in any discrete algorithm that aims to exploit the smoothness of the square norm $\normDen{Aa-\obs}^2$ in 
		\eqref{eq:InvDif:Regularised}. 
		Given a discretization scheme, these finite-dimensional approximations are the operators
		$\widetilde{A}:\ASpace_{q_1}\rightarrow \DenSpace_{q_2}$ and $\widetilde{A^*}:\DenSpace_{q_2}\rightarrow \ASpace_{q_1}$ 
		such that~\cite[p.~964]{Zeidler1990}
		\begin{IEEEeqnarray}{rl}  \IEEEyesnumber \label{eq:discopersformula} \IEEEyessubnumber* \label{eq:discopersformula:1}
			\widetilde{A} \tilde{a}\, &= \Restr{\DenSpace}\left( A \Exten{\ASpace}\left[\tilde{a}\right] \right),\,\forall \tilde{a} \in \ASpace_{q_1}\,,\\ \label{eq:discopersformula:2}
			\widetilde{A^*} \tilde{d}\, &= \Restr{\ASpace}\left( A^* \Exten{\DenSpace}\left[\tilde{d}\right] \right),\,\forall \tilde{d} \in \DenSpace_{q_2}\,.			
		\end{IEEEeqnarray}
		Similarly, any operators $B_1:\ASpace \rightarrow \ASpace$, $B_2:\DenSpace \rightarrow \DenSpace$ will be approximated within the 
		discretization scheme by $\widetilde{B}_1:\ASpace_{q_1} \rightarrow \ASpace_{q_1}$ and 
		$\widetilde{B}_2:\DenSpace_{q_2} \rightarrow \DenSpace_{q_2}$ such that 
		\begin{IEEEeqnarray}{rl} \IEEEyesnumber \label{eq:discselfopersformula}\IEEEyessubnumber* \label{eq:discselfopersformula:1}
			\widetilde{B}_1 \tilde{a}\, & = \Restr{\ASpace}\left(B_1 \Exten{\ASpace}\left[ \tilde{a} \right] \right),\,\forall \tilde{a} \in \ASpace_{q_1}\,, \\ \label{eq:discselfopersformula:2}
			\widetilde{B}_2 \tilde{d} \, & = \Restr{\DenSpace}\left(B_2 \Exten{\DenSpace}\left[ \tilde{d} \right] \right),\,\forall \tilde{d} \in \DenSpace_{q_2}\,,
		\end{IEEEeqnarray}
		and any functional $\vartheta:\ASpace\rightarrow \reals$ will be approximated by $\tilde{\vartheta}:\ASpace_{q_1} \rightarrow \reals$
		such that
		\begin{IEEEeqnarray}{c} \label{eq:discfunctionalsformula}
			\tilde{\vartheta}\left(\tilde{a}\right) = \vartheta\left(\Exten{\ASpace}\left[ \tilde{a} \right] \right),\, \forall \tilde{a} \in \ASpace_{q_1}\,.
		\end{IEEEeqnarray}
	
		In our particular case, we have chosen the following restriction and extension operators, and through them, a specific discretization. 
			The restriction operator for $\DenSpace$ is given by the camera and fulfills \eqref{eq:Rd}.  
			For restricting $\ASpace$, then, we propose using 
			\begin{IEEEeqnarray}{c} \label{eq:Ra}
				\Restr{\ASpace}\left(a\right) = \left\lgroup \tilde{a}_{m,n,k} \right\rgroup \!
				 = \! \left\lgroup \frac{1}{\sqrt{\Delta_k}}\int_{\discreg{m,n,k}} \!\!\!\!\!\!\!\! a(\pos,\sigma)\,\dint\pos \dint\sigma \right\rgroup\!,
			\end{IEEEeqnarray}
			with $\ASpace_{q_1}=\matrices{M,N,K}$, $m,n$ as above, $k\in\listint{K}$,
			$\discreg{m,n,k} =\discreg{m,n} \times \left[ \dsigma_{k-1}, \dsigma_k \right]$ and $ \Delta_k = \left(\dsigma_k - \dsigma_{k-1} \right)$,
			with $\lbrace \dsigma_0, \dsigma_1, \dots, \dsigma_K\rbrace$ an arbitrary grid in the $\dsigma$-dimension 
			such that $\tilde{\sigma_{k-1}} < \dsigma_k$, $\dsigma_0=0$ and $\dsigma_K=\dsigmax$. As mentioned before, this discretization
			is also assumed in user parameters that concern $a(\pos,\sigma)$, i.e., $\tilde{\mu}\in\matrices{M,N}$ is considered as a mask in the 
			finite-dimensional spatial coordinates and $\tilde{\xi}\in\reals_+^K$ is considered as a non-negative weighting vector across
			$k$s. For the latter, note that if $\xi$ has the structure discussed at the end of Section~\ref{sec:InvDif:Opti}, i.e., it is the 
			$(0,1)$-indicator
			of a set $\aleph\subset[0,\sigmax]$, this set will be aligned with respect to the discretization boundaries $\dsigma_k$, and an equivalent
			set of discrete indexes 
			$\tilde{\aleph}=\lbrace k \in \listint{K} : (\dsigma_{k-1},\dsigma_k)\subset\aleph \rbrace$ can be defined.
			
			For the extension operators, we use the inner approximation interpretation of the discrete spaces 
			$\ASpace_{q_1}\subset \ASpace$ and $\DenSpace{q_2}\subset \DenSpace$, and consider them as parameterizations of piece-wise
			constant functions in $\ASpace$ and $\DenSpace$, i.e.
			\begin{IEEEeqnarray}{c} \label{eq:Ed}
				\Exten{\DenSpace}\left(\tilde{d}\right) 
				= \sum_{n=1}^N \sum_{m=1}^{M} \tilde{d}_{m,n} i_{\discreg{m,n}}  \,,
			\end{IEEEeqnarray}
			and
			\begin{IEEEeqnarray}{c} \label{eq:Ea}
				\Exten{\ASpace}\left(\tilde{a}\right) 
				= \sum_{n=1}^N \sum_{m=1}^{M} \sum_{k=1}^{K} \frac{1}{\sqrt{\Delta_k}} \tilde{a}_{m,n,k} \, i_{\discreg{m,n,k}}  \,,
			\end{IEEEeqnarray}
			with $i_{\discreg{m,n}}$ and $i_{\discreg{m,n,k}}$ the $(0,1)$-indicator functions for 
			$\discreg{m,n}$ and $\discreg{m,n,k}$, respectively.
			
\begin{figure}
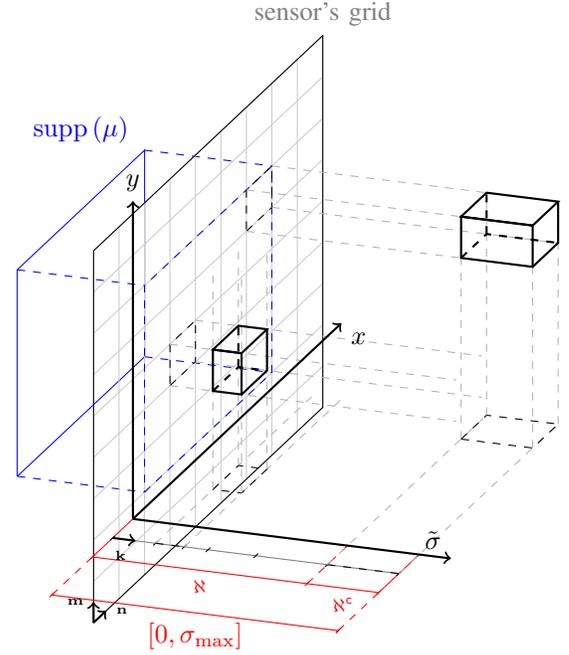

			\centering
				\def\sca{.9}
					\ifundef{\sca}{\def\sca{1.8}}{}
	\tdplotsetmaincoords{70}{110}
	\begin{tikzpicture}[tdplot_main_coords,scale=\sca]
		
			\def\xdist{3.5} \def\sigminn{0} \def\sigmaxn{4.5}
			\def\sigdist{-2} \def\xmin{-.5} \def\xmax{-6} \def\ymin{0.25} \def\ymax{3.5}
			\draw[blue] plot coordinates 
			{(\xmin,\sigdist,\ymin) (\xmax,\sigdist,\ymin) (\xmax,\sigdist,\ymax) (\xmin,\sigdist,\ymax) 
			 (\xmin,\sigdist,\ymin)};
			\node[color=blue] at (\xmax,\sigdist-1,\ymax+.2) {$\supp{\mu}$};
			\foreach \x in {\xmin,\xmax} {
				\foreach \y in {\ymin,\ymax}{
					\draw[thin,dashed,blue] (\x,\sigdist,\y) -- (\x,0,\y);
				}
			}
			
			\def\nbins{5} \def\binsextra{2} \pgfmathsetmacro{\nbinsextrapo}{-\binsextra +1} \pgfmathsetmacro{\nbinsextra}{\nbins +\binsextra}
			\pgfmathsetmacro{\xbynbins}{(\xmax-\xmin)/\nbins} \pgfmathsetmacro{\ybynbins}{(\ymax-\ymin)/\nbins}
			\foreach \n in {-\binsextra, \nbinsextrapo, ...,\nbinsextra}{
				\draw[gray!40,thin] (\xmin + \n*\xbynbins,0,\ymin-\binsextra*\ybynbins) -- (\xmin + \n*\xbynbins,0,\ymin+\nbinsextra*\ybynbins);
				\draw[gray!40,thin] (\xmin-\binsextra*\xbynbins,0,\ymin + \n*\ybynbins) -- (\xmin+\nbinsextra*\xbynbins,0,\ymin + \n*\ybynbins);
			}
			\draw[thin] (\xmin -\binsextra*\xbynbins,0,\ymin-\binsextra*\ybynbins) -- 
						(\xmin +\nbinsextra*\xbynbins,0,\ymin-\binsextra*\ybynbins) --
						(\xmin +\nbinsextra*\xbynbins,0,\ymin+\nbinsextra*\ybynbins) node[anchor=south] {\textcolor{gray}{sensor's grid}} --
						(\xmin -\binsextra*\xbynbins,0,\ymin+\nbinsextra*\ybynbins) -- cycle;
			\draw[blue,thin,dashed] plot coordinates 
			{(\xmin,0,\ymin) (\xmax,0,\ymin) (\xmax,0,\ymax) (\xmin,0,\ymax) 
			 (\xmin,0,\ymin)};
			\draw[red] (.1+\xdist,0,0) -- (-.1+\xdist,0,0) -- 
				   (\xdist,\sigminn,0)  -- node[midway,below]{$\left[0,\sigmax\right]$} (\xdist,\sigmaxn,0) -- 
				   (.1+\xdist,\sigmaxn,0) -- (-.1+\xdist,\sigmaxn,0); 
			\draw[thin,dashed,red] (\xdist,0,0) -- (0,0,0);
			\draw[thin,dashed,red] (\xdist,4.5,0) -- (0,4.5,0);
			\draw[gray,thin] (.25*\xdist,\sigminn,0) -- (.25*\xdist,\sigmaxn,0);
 			\pgfmathsetmacro{\lofsig}{\sigmaxn-\sigminn}
 			\foreach \prop in {0, .15, .25, .333, .5, .75, 1}{
 				\draw (.1+.25*\xdist,\prop*\lofsig,0) -- (-.1+.25*\xdist,\prop*\lofsig,0);
 			}
 			\draw[red,thin] (.5*\xdist,\sigminn,0) -- node[midway,below]{\scriptsize $\aleph$} (.5*\xdist,\sigminn+.75*\lofsig,0);
 			\draw[red,thin] (.5*\xdist,\sigminn+.75*\lofsig,0) -- node[midway,below]{\scriptsize $\complement{\aleph}$} (.5*\xdist,\sigmaxn,0);
 			\foreach \prop in {0, .75, 1}{
 				\draw[red,thin] (.1+.5*\xdist,\prop*\lofsig,0) -- (-.1+.5*\xdist,\prop*\lofsig,0);
 				\draw[red,thin,dashed] (.5*\xdist,\prop*\lofsig,0) -- (0,\prop*\lofsig,0);
 			}
		
		\draw[thick,->] (\xmin-\binsextra*\xbynbins,0,\ymin-\binsextra*\ybynbins) 
					 -- (\xmin-\binsextra*\xbynbins+ .5*\xbynbins,0,\ymin-\binsextra*\ybynbins) 
						node[anchor=west]{\tiny $\mathbf{n}$};
		\draw[thick,->] (\xmin-\binsextra*\xbynbins,0,\ymin-\binsextra*\ybynbins) 
					 -- (\xmin-\binsextra*\xbynbins, 0, \ymin+ .5*\ybynbins-\binsextra*\ybynbins) 
						node[anchor=east]{\tiny $\mathbf{m}$};
		\draw[thick,->] (.25*\xdist,0,0) -- (.25*\xdist,.08*\lofsig,0) node[anchor=north east]{\tiny $\mathbf{k}$};
		
		\def\m{2} \def\n{1} \def\sigselmin{.15} \def\sigselmax{.25} 
		\input{\figs/discretization_reg}
		\def\m{4} \def\n{4} \def\sigselmin{.75} \def\sigselmax{1} 
		\input{\figs/discretization_reg}
		
		\draw[thick,->] (0,0,0) -- (-9,0,0) node[anchor=north west]{$x$};
		\draw[thick,->] (0,0,0) -- (0,5,0) node[anchor=south east]{$\dsigma$};
		\draw[thick,->] (0,0,0) -- (0,0,5) node[anchor=south]{$y$};
	\end{tikzpicture}
				\vspace{-5pt}
			\caption{Example of a discretization grid for $\ASpace$ with $M=N=9$, $K=6$. Highlighted, $\discreg{5,4,2}$ and 
					$\discreg{7,7,6}$. In gray, is the sensor's grid, which coincides with the spatial grid for $\ASpace$ and the resolution
					of the recovered PSDR. In blue, is the support of a mask $\mu(\pos)$ that specifies where particle sources
					can be located. Note that the particular support can also be specified in terms of a mask matrix $\tilde{\mu}$.
					In red, are the sets $\aleph$ and $\complement{\aleph}$ that characterize the behavior of $\xi$. Note that here,
					$\tilde{\aleph}=\left\lbrace 1,2,\dots,5\right\rbrace$ and $\complement{\tilde{\aleph}} = \lbrace 6 \rbrace$.
					 \label{fig:discgridPSDR}}
		\end{figure}

		An example of the structure chosen for the discretization of $\ASpace$ is portrayed in Fig.~\ref{fig:discgridPSDR}. 
		Because the camera's restriction operator \eqref{eq:Rd} fixed the understanding of the spatial domain in $\DenSpace$ as a regular
		grid, it was natural to use the same regular grid structure for the spatial dimension in the discretization of $\ASpace$ too.
		The discretization of the $\sigma$-dimension, however, could have been addressed much differently, for example, using a more flexible
		function basis in \eqref{eq:Ea}. However, we opted to use an irregular grid in $\sigma$, which provides modeling flexibility and preserves mathematical
		tractability.
		
		In terms of the dimensionality of the problem, our choices imply that $\ASpace$ is discretized with the same spatial resolution as 
		the observation $\dobs$. A finer resolution in this discretization would yield super-resolution in the recovery of the PSDR and, 
		therefore, more accurate SL. However, even with the modest typical values $M=N=512$ and $K=8$ used in our numerical evaluations in 
		Part~II~\cite{AguilaPla2017a}, our choice already results in a discretized inverse problem with $q_1 \approx 2\cdot10^6$ optimization 
		variables. In real scenarios, like the one introduced in Section~\ref{sec:NumRes:Real}, these values are $M=N=2048$ and $K=6$, which 
		result in $q_1\approx25\cdot10^6$ unless the resolution of the sensor is artificially decreased. Therefore, we have left further 
		inquiries into grid-based super-resolution methods outside of the scope of this paper.
				
		Direct computation of \hyperref[eq:discopersformula:1]{(\ref*{eq:discopersformula})} using the particular restriction and extension operators 
		\eqref{eq:Rd}, \eqref{eq:Ra}, \eqref{eq:Ed}, and \eqref{eq:Ea} yields
		\begin{IEEEeqnarray}{rl}
			\widetilde{A} \tilde{a} & =  \sum_{k=1}^{K} \tilde{g}_k \circledast \tilde{a}_k,  \label{eq:discdiff} \\
			\widetilde{A^*} \tilde{d} & = \left\lgroup 
											\tilde{\mu} \odot \left(\tilde{g}_k \circledast \left[\tilde{w}^2 \odot \tilde{d}\right] \right) 
										  \right\rgroup\,, \label{eq:discadj}
		\end{IEEEeqnarray}
		for $k \in\listint{K}$, where $\cdot^2$ refers to element-wise squaring, $\odot$ to the Hadamard (element-wise) product, 
		$\circledast$ to discrete convolution with zero-padding, $\tilde{a}_k\in\matrices{M,N}$ for $k\in\listint{K}$ to the different cuts in 
		the $k$-dimension in $\tilde{a}$, and $g_k:\integ^2\rightarrow \reals_+$ for $k\in\listint{K}$ to the discrete convolutional kernels
		such that
		\begin{IEEEeqnarray*}{c}
			\tilde{g}_k(\tilde{\pos}) = \frac{1}{\sqrt{\Delta_k}} \int_{\dsigma_{k-1}}^{\dsigma_k} \int_{\discreg{0,0}^2} 
								g_{\dsigma}\left( \tilde{\pos} + \boldsymbol{\rho}_1 - \boldsymbol{\rho}_2 \right) 
								\dint\boldsymbol{\rho}_1 \times \dint\boldsymbol{\rho}_2 \dint \dsigma\,,
		\end{IEEEeqnarray*}
		for any $\tilde{\pos}\in\integ^2$. 
		Note that, because $g_{\dsigma}$ is an isotropic 2D Gaussian probability density function, it is separable in the different spatial 
		dimensions, and, therefore 
		\begin{IEEEeqnarray}{c} \label{eq:gk}
			\tilde{g}_k\!\left[ (m, n) \right] = \frac{1}{\sqrt{\Delta_k}}\int_{\dsigma_{k-1}}^{\dsigma_k} \omega_{\dsigma}(m) \omega_{\dsigma}(n) 
			\dint\dsigma\,,
		\end{IEEEeqnarray}
		$\forall (m,n) \in\integ^2$ with $\omega_{\dsigma}:\integ\rightarrow\reals_+$ such that for any $m\in\integ$, \cite{Owen1980}
		\begin{IEEEeqnarray*}{c}
			\omega_{\dsigma}(m) = \int_{-\frac{1}{2}}^{\frac{1}{2}} \left[\Phi\!\left(\frac{m+\rho+\frac{1}{2}}{\dsigma}\right) - 
																	   \Phi\!\left(\frac{m+\rho-\frac{1}{2}}{\dsigma}\right) \right] \dint \rho\,,
		\end{IEEEeqnarray*}
		where $\Phi:\reals\rightarrow[0,1]$ is the standard 1D normal cumulative density function.
		The detailed derivation of the expressions \eqref{eq:discdiff}, \eqref{eq:discadj} and \eqref{eq:gk}, as well as insights on techniques for the
		numerical computation of \eqref{eq:gk}, can be found in the supplementary material to this paper.
	
	Using the proposed discretization scheme we obtain a discretized equivalent to \eqref{eq:InvDif:Regularised}, i.e., the  
	finite-dimensional optimization problem
	\begin{equation} \label{eq:optidisc}
			\min_{\tilde{a}}
			\left\lbrace 
				\left\| \tilde{A}\tilde{a} - \dobs \right\|_{\tilde{w}}^2
				 + \lambda \sum_{m,n} \! \left\| \tilde{\xi} \odot \tilde{a}_{m,n} \right\|_2
			\right\rbrace,
	\end{equation}
	subject to $\tilde{a} \in \matricesP{M,N,K}$, where $\| \cdot \|_{\tilde{w}}$ denotes the finite-dimensional $\tilde{w}$-weighted Euclidean norm, 
	and $\tilde{a}_{m,n}\in\reals^K$. Now, the finite dimensionality of the problem enables us to rely on the extreme-value
	theorem for deriving the existence of a minimizer of \eqref{eq:optidisc} from the closed and bounded sublevel sets given by the regularizer 
	when $\lambda > 0$ and $\tilde{\xi}_k>0$ for any $k$, or from those of the data penalty term if $\tilde{\xi}_k=0$ for some $k$ or $\lambda=0$.
	This is summarized in Lemma~\ref{lem:existence}.
	
	\begin{lemma}[Existence of a solution to the discretized problem]\label{lem:existence}
		Consider the function $C:\matrices{M,N,K} \rightarrow \bar{\reals}$ such that $C(\tilde{a}) =\left\| \dobs - \tilde{A}\tilde{a} \right\|_{\tilde{w}}^2 + f(\tilde{a})$,
		with $f:\matrices{M,N,K} \rightarrow \bar{\reals}$ such that
		\begin{IEEEeqnarray*}{c}
			f(\tilde{a}) = \delta_{\matricesP{M,N,K}}(\tilde{a}) + \lambda \sum_{m,n} \! \left\| \tilde{\xi} \odot \tilde{a}_{m,n} \right\|_2\,.
		\end{IEEEeqnarray*}
		Then, if either $\lambda>0$ and $\tilde{\xi}_k>0$ for any $k$, or $\tilde{w}_{m,n}>0$ for any $(m,n)$, $\exists \tilde{a}_\mathrm{opt}\in \matricesP{M,N,K}$ such that
		$C(\tilde{a}_\mathrm{opt}) = \inf_{\tilde{a}\in\matrices{M,N,K}} C(\tilde{a})$, i.e. \eqref{eq:optidisc} has a minimizer.
	\end{lemma}
	The extension of Lemma~\ref{lem:existence} to function spaces is challenging even in the case in which $\lambda>0$ and $\xi(\sigma)>0~\ae$ in $[0,\sigmax]$.
	Even if it was possible to extract closed and bounded sublevel sets from the behavior of the regularizer, the characterization of compact sets in $\mathrm{L}^p$ involves 
	$\mathrm{L}^p$-equicontinuity, which does not seem to follow easily for our problem set-up. Nonetheless, the case $p=2$ is slightly more tractable \cite{Pego1985}, and the 
	possibility remains that the smoothing property of the kernels that compose the diffusion operator $A$ can somehow be exploited. In any case, further characterizing the
	diffusion operator $A$, its range, nullspace and spectrum would surely help in addressing this issue and understanding its specifics.
	Uniqueness statements are challenging to obtain for both the continuous and discrete formulations. However, the intuition remains that, by coupling the 
	third dimension with the non-negative group-sparsity regularizer, the optimization does not only get biased towards more plausible explanations of the data in terms of 
	stationary sources, but the inverse problem also improves its condition by treating differently different $a\in\ASpace_+$ that approximate the observation at the same
	level of accuracy $\normDen{Aa-\obs}$.

		\section{Example on Real Data and Closing Remarks} \label{sec:Conclusions}

		\begin{figure*} 
			\centering
			\includegraphics[width=.493\linewidth,keepaspectratio=true]{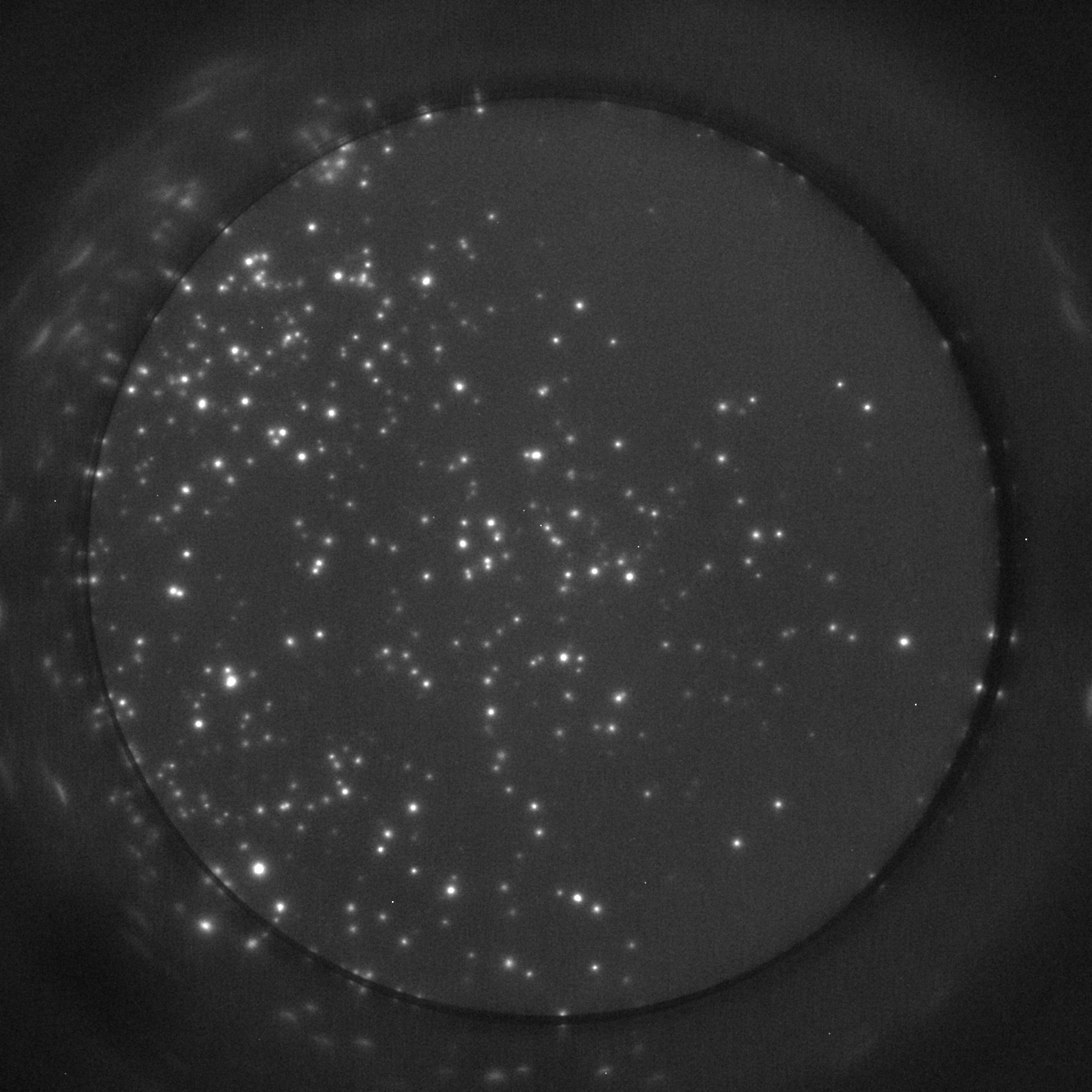}
			\includegraphics[width=.493\linewidth,keepaspectratio=true]{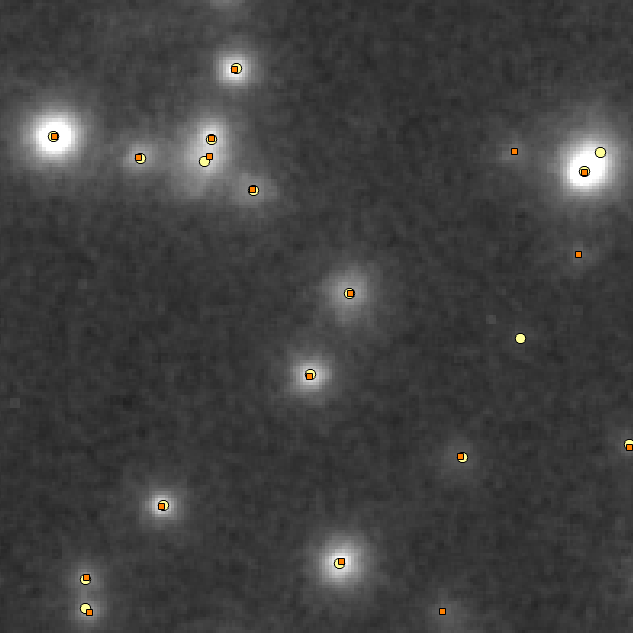}
			\caption{ \label{fig:dataexample}
				Example of SL performance on a section of real Fluorospot data.
				To the left, grayscale image recovered from the raw RGB data, with increased 
				luminosity. To the right, detection results (yellow circles) and human labeling 
				(orange squares) for a specific section displayed on top of the grayscale image with increased luminosity.
			}
		\end{figure*}
	\subsection{Example on real data} \label{sec:NumRes:Real}

		In Part II of this paper \cite{AguilaPla2017a} we provide, along with the algorithmic developments, an extensive quantitative assessment of 
		the proposed SL methodology. However, in order to keep this Part I self contained and exhibit the benefits of the contributed modeling and inverse problems
		framework, we analyzed a real Fluorospot image and compared the results to expert human labeling. In particular, we discretized an algorithm to solve 
		\eqref{eq:InvDif:Regularised} with $\lambda = 4000$, using $K=6$ and $\lbrace \dsigma_0,\dsigma_1,\dots,\dsigma_6\rbrace = \lbrace 2, 15, 20, 30, 40, 50, 
		70 \rbrace$. The details and approximations in the algorithmic solution were analogous to those used in the numerical results of
		Part~II~\cite[Section III]{AguilaPla2017a}. This resulted in a discretized recovered PSDR $\tilde{a}_{\mathrm{opt}}$ that, after minor post-processing 
		(see Part~II~\cite[Section III]{AguilaPla2017a}) yielded an F1-Score relative to the human 
		labeling of $0.9$, with precision $0.92$ and recall $0.88$. A visualization of the image and some of the SL results are shown in Fig.~\ref{fig:dataexample}. 
		In terms of the cell count, our algorithm obtained $346$ cell locations, while the human labeling contained $360$ locations.
		
		This image was obtained from a biochemical assay in which FITC dye was used as a marker, and it was captured by an RGB sensor that produced 
		raw data with dimensions $M=N=2048$ and a dynamic range of $[0,2^{16}-1]$. This raw data was subject to a Bayer color filter array \cite{Lukac2008}, in which neighboring pixels 
		correspond to different color bands. Because the different color filters have different sensitivities to the particular wavelength
		emitted by the FITC molecules, neighboring pixels were weighted accordingly to estimate the underlying luminosity.
		The weighting function $\tilde{w}(\tilde{\pos})$ was also updated to weight the errors in each position with respect to
		the sensor's sensitivity in that position. Additionally, the area comprised by the well was selected manually, and any position
		outside it was given weight zero, i.e. $\tilde{w}(\tilde{\pos})=0$.
		Finally, the mask function $\mu(\tilde{\pos})$ was set to $1$ for all $\tilde{\pos}\in\integ^2$.
		
		In our opinion, the SL results produced by our algorithm are of a quality comparable to that of the expert human labeling. 
		Indeed, in many cases the recovered location seems to be more reliable, and the criteria to determine what constitutes a true detection
		seems to be more consistent throughout the image. 
	
	\subsection{Closing remarks}
	
		In this first part of our paper, we have developed a novel observation model for images that measure
		3D reaction-diffusion-adsorption-desorption physical processes. This model provides an objective means to generate reliable synthetic data
		for biochemical assays from the specific physical parameters that characterize them.
		We have then proposed an optimization framework to recover reaction, i.e. particle secretion, by exploiting the assumption that it was spatially localized
		and temporally continuous. These properties are common in the context of SL in diffusion models, and are easy to interpret from the perspective of 
		biochemical assays. Moreover, the designed optimization framework allows for the inclusion of different kinds of prior information, which can impact
		practical use greatly. To finalize the paper, we have introduced a simple discretization scheme to implement both synthesis and 
		analysis methods based on our model, and we have provided some experimental results on real Fluorospot data.
		
		In Part II of our paper \cite{AguilaPla2017a}, we develop an accelerated proximal gradient algorithm to solve the functional 
		optimization problem in \eqref{eq:InvDif:Regularised}, providing an expression for the proximal operator of the non-negative 
		group-sparsity regularizer. We also use the discretization scheme we presented here to derive 
		an efficient implementation that approximates solutions of \eqref{eq:optidisc}. Finally, we provide thorough empirical evaluation of our algorithm
		both in terms of detection and in terms of optimal transport metrics.


		
		\appendices
			\section{Characterization of the model from physical parameters} \label{app:PhyPar}

	
	As in Section~\ref{sec:DataModel:ObsModel}, we will interchangeably rely on macroscopic arguments pertaining to the evolution of particle distributions, governed by \eqref{eq:pde}, 
	and microscopic arguments pertaining to the behavior of individual particles \cite{Ursell2007}.
	We start by presenting the proof of the characterization of the model in a simplified case, i.e., $\kd = 0$.
	\begin{IEEEproof}[Proof - Lemma~\ref{lem:charvarphi-noescape} (Characterization of the observation model from physical parameters, Case $\kd=0$)]
		From \cite[Equation (3.1)]{Singer2008} or \cite[Equations (10), (22) and (27)]{Agmon1984} we have that, particles released at
		time $0$ are, at time $t$, distributed in the $z$-dimension according to the density 
		\begin{IEEEeqnarray*}{rl} 
			u(z,t) & = \frac{1}{\sqrt{\pi D t}} \exp\left( -\frac{z^2}{4Dt} \right) \\
			& {-}\: \frac{\ka}{D} \exp\left( \frac{\ka z+\ka^2 t}{D} \right) \mathrm{erfc}\left( \frac{z}{\sqrt{4Dt}} + \ka \sqrt{\frac{t}{D}} \right) \nonumber \, ,
		\end{IEEEeqnarray*}
		for $z\geq 0$. For $\kd=0$, bound particles are never released. Thus, the time until the first binding event is the same as the total time in free motion.
		For a particle released at time $0$, the probability density function of the time until the first binding event is \cite{Berezhkovskii2004}
		$\phi(\tau) = \ka u(0,\tau)$. Because $\tau$ was defined as the total time in free motion within the time window $[0,t)$, we have that
		$\varphi(\tau,t) = \phi(\tau) i_{[0,t)}(\tau)$.
	\end{IEEEproof}
	We now present how this result is extended to $\kd\geq 0$.
	\begin{IEEEproof}[Proof - Theorem~\ref{th:charvarphi-general} (Characterization of the observation model from physical parameters)]
		Consider first that, for $\kd=0$, \eqref{eq:sum-varphi} particularizes to \eqref{eq:first-adsorption}, and, thus, our statement is already proved in 
		Lemma~\ref{lem:charvarphi-noescape}. 
		
		Consider then the case $\kd > 0$. 
		Then, for any given time window, a particle that at the end of that period is bound has some probability of having been bound and remained still 
		thereafter; some probability of having been bound, disassociated, and then bound again; some probability of having been bound and disassociated twice, and 
		then bound again, and so on. Note that we focus only on those particles that are found bound at the end of a specific period, as those are 
		the ones modeled by $\varphi(\tau,t)$ and \eqref{eq:observation-model}. For any $i\in\lbrace1,2,\dots\rbrace$, consider the random variables $\tau_i$ 
		and $\eta_i$, that represent the time spent in free motion and bound, respectively, the $i$-th time a particle goes through this cycle.
		The time invariance of \eqref{eq:pde}, i.e., the lack of memory in the diffusion, association and disassociation processes, yields that 
		$\{ \tau_i \}_{i=1}^\infty$ and $\{ \eta_i \}_{i=1}^\infty$ are mutually independent sequences of independent and equally distributed 
		random variables. 
		
		Because both after release and after disassociation particles start their free motion at the surface, i.e. $z=0$,
		the distribution of any specific $\tau_i$ is given by $\phi(\tau_i)$ as in \eqref{eq:first-adsorption}, Lemma~\ref{lem:charvarphi-noescape}.
		Second, the probability density $\psi(\eta_i)$ of any $\eta_i$ may be obtained from 
		$\frac{\partial}{\partial \eta_i} \psi(\eta_i) = -\kd \psi(\eta_i)$ [cf. \eqref{eq:pde-surface}] and $\int_{\eta=0}^\infty \psi(\eta_i) \dint\eta_i = 1$ 
		as $\psi(\eta_i) = \kd^{-1}e^{-\kd \eta_i}$. 
		
		Consider now $\tau^{(j)}$, a random variable representing the total amount of time in free motion before the $j$-th disassociation event.
		Then, $\tau^{(j)} = \sum_{i=1}^{j} \tau_i$. 
		Because $\tau_1,\tau_2,\dots,\tau_j$ are independent and identically distributed with density $\phi(\tau_i)$, the density of $\tau^{(j)}$ is given 
		by the $j$-th convolutional power $\phi^{j*}(\tau^{(j)})$ of $\phi(\tau)$.
		Similarly, consider $\eta^{(j)}$, a random variable representing the total time a particle has been bound to the surface before the $j$-th
		disassociation event. Then, $\eta^{(j)} = \sum_{i=1}^{j} \eta_i$ and is distributed with density $\psi^{j*}(\eta^{(j)})$.
		
		We now obtain the probability that a particle released at time $0$ remains bound 
		at time $t$, after having been bound a total of exactly $j$ times, and having spent a total time
		$\tau=\tau^{(j)}$ in free motion. Note that this is equivalent to stating that $\eta^{(j)} \geq t-\tau^{(j)}$. Therefore,
		\begin{IEEEeqnarray*}{rl}
			\Pr\left(\tau = \tau^{(j)}\right) & = \Pr\left(\eta^{(j)} \geq t-\tau^{(j)} \right) 
			 = \int_{t-\tau^{(j)}}^{+\infty} \psi^{j*}(\eta) \dint\eta \\& = p\left[j-1;\kd\left(t-\tau^{(j)}\right)\right]\,,
		\end{IEEEeqnarray*}
		where $p[\cdot;\cdot]$ is the Poisson probability mass function as in \eqref{eq:Poisson}. This result is 
		naturally expected, as it merely states that the probability of the sum of the first $j$ binding times 
		exceeding $t-\tau^{(j)}$ is the same as the probability of having exactly $j-1$ disassociation events 
		during a time span of $t-\tau^{(j)}$. Indeed, at a disassociation rate of $\kd$, this probability is
		given by the Poisson distribution with parameter $\lambda = \kd\left(t-\tau^{(j)}\right)$.

		Using the law of total probability and that $\tau$ was defined as the time in free motion
		before time $t$ only, we have that		
		\begin{IEEEeqnarray*}{rl}
			\varphi(\tau,t) & = i_{[0,t)}(\tau) \sum_{j=1}^\infty \phi^{j*}(\tau) \Pr\left(\tau = \tau^{(j)}\right) \\
			&= i_{[0,t)}(\tau) \sum_{j=1}^\infty \phi^{j*}(\tau) p\left[j-1;\kd(t-\tau)\right] \,.
		\end{IEEEeqnarray*} 
	\end{IEEEproof}
	Finally, we present the proof Lemma~\ref{lem:Truncate},
	that provides the truncation point of the infinite sum in Theorem~\ref{th:charvarphi-general} at which a certain
	accuracy $\epsilon>0$ is guaranteed.
	\begin{IEEEproof}[Proof - Lemma~\ref{lem:Truncate} (Truncation of the sum to characterize the model)]
		Note first that, $\forall \lambda_1,\lambda_2>0$ such that $\lambda_1>\lambda_2$, we have that 
		$\sum_{j=J}^{+\infty}p[j,\lambda_1] > \sum_{j=J}^{+\infty}p[j,\lambda_2]$ for any $J\in\natu$.
		Now, consider that, because $\phi(\tau)$ is a probability density defined for $\tau\in[0,+\infty)$, if we consider its extension by zeros,
		$\phi:\reals\rightarrow\reals_+$ such that $\phi(\tau)=0$ for $\tau<0$, we have that 
		$\|\phi\|_{\mathrm{L}^1\left(\reals\right)}=1$. Then, using Young's inequality (see \cite{Beckner1975} for details), we have that 
		$\normLebTwo{\reals}{\phi^{(k+1)*}} \leq \normLebTwo{\reals}{\phi^{k*}}$ for $k\geq 1$ and thus, 
		$\normLebTwo{\reals}{\phi^{k*}} \leq \normLebTwo{\reals}{\phi} = \normLebTwo{0,+\infty}{\phi}$ for any $k\geq1$.

		Using Young's inequality again \cite{Beckner1975} but with $r=\infty$, $p=q=2$ and $n=1$, we obtain 
		that for any $f,g\in\LebTwo{\reals}$,
		$\left\|f * g\right\|_{\mathrm{L}^{\infty}\left(\reals\right)} \leq \normLebTwo{\reals}{f} \normLebTwo{\reals}{g}$.
		
		Then, we have that for any $(\tau,t)\in[0,T]^2$ such that $\tau < T$,
		\begin{IEEEeqnarray*}{rl}
			\tilde{\varphi}(\tau,t) & = \sum_{j=J_\epsilon}^{+\infty} \phi^{j*}(\tau) p\left[j-1;\kd(t-\tau)\right] \\
			& \leq \sum_{j=J_\epsilon}^{+\infty} \left\|\phi^{j*}\right\|_{\mathrm{L}^{\infty}\left(\reals\right)} p\left[j-1;\kd(t-\tau)\right] \\
			& \leq \sum_{j=J_\epsilon}^{+\infty} \normLebTwo{\reals}{\phi} \normLebTwo{\reals}{\phi^{(j-1)*}} p\left[j-1;\kd(t-\tau)\right] \\
			& \leq \normLebTwo{0,+\infty}{\phi}^2 \sum_{j=J_\epsilon}^{+\infty}  p\left[j-1;\kd(t-\tau)\right] \\
			& \leq  \normLebTwo{0,+\infty}{\phi}^2 \sum_{j=J_\epsilon}^{+\infty}  p\left[j-1;\kd T\right] \\
			& \leq \normLebTwo{0,+\infty}{\phi}^2 \frac{\epsilon}{\normLebTwo{0,+\infty}{\phi}^2} = \epsilon\,.
		\end{IEEEeqnarray*} \vspace{-12pt}
	\end{IEEEproof}

			\section{Diffusion and Inverse diffusion} \label{app:DiffOp}

	In this appendix, we present the proof to Lemmas~\ref{lem:NormDiff} to \ref{lem:AdjDiff}, which characterize the diffusion operator
	$A$ in Definition~\ref{def:DiffOp}. Furthermore, we provide a proof of Lemma~\ref{lem:existence}, that guarantees the existence of a
	solution for the discretized version of the optimization problem that we propose for inverse diffusion.
	
	Consider first Properties \ref{prop:NormOneGauss} and \ref{prop:SelfGaus}, which constitute a characterization
	of the Gaussian blur operators $\left\lbrace G_\sigma \right\rbrace_{\sigma>0}$ in Definition~\ref{def:GaussBlur} in terms of their norm and adjoint operators.
	\begin{property}[Norm of the Gaussian blur operator] \label{prop:NormOneGauss}
		The Gaussian blur operators in Definition~\ref{def:GaussBlur} have norm $1$, i.e.,
		\begin{IEEEeqnarray*}{c}
			\left\|G_\sigma\right\|_{\opers{\LebTwo{\reals^2}}{\LebTwo{\reals^2}}} = \normLebOne{\reals^2}{g_\sigma} = 1\,.
		\end{IEEEeqnarray*}
	\end{property}
	\begin{IEEEproof}
		For any $f\in\LebTwo{\reals^2}$, let $\check{f}\in\LebTwo{\reals^2}$ denote its Fourier transform. Then,
		we have that
		\begin{IEEEeqnarray}{rl} \IEEEyesnumber \label{eq:normGauss} \IEEEyessubnumber* \label{eq:normGauss:first}
			\normLebTwo{\reals^2}{f * g_\sigma} & = \normLebTwo{\reals^2}{\check{f}\check{g}_\sigma} \leq |\check{g}_\sigma(\mathbf{0})| \normLebTwo{\reals^2}{\check{f}} \\
			& = \int_{\reals^2} g_\sigma(\boldsymbol{\rho}) \dint\boldsymbol{\rho} \normLebTwo{\reals^2}{\check{f}} \label{eq:normGauss:second} \\
			& = \normLebOne{\reals^2}{g_\sigma} \normLebTwo{\reals^2}{f}\,. \nonumber
		\end{IEEEeqnarray}		
		In \eqref{eq:normGauss:first}, we have used the Parseval-Plancherel theorem and the convolution theorem followed by the bound on the Fourier transform of any real and non-negative function
		by the value at its origin. In \eqref{eq:normGauss:second}, we have used again the Parseval-Plancherel theorem, the expression for the Fourier transform 
		evaluated at $\mathbf{0}$, and the non-negativity of the Gaussian kernel.
		
		Consider now the sequence of Gaussian kernels with standard deviation $n\in\natu$, i.e., $\lbrace g_n \rbrace_\natu \subset\LebTwo{\reals^2}$ and observe that 
		$\normLebTwo{\reals^2}{g_n} = 1/(2\pi n)$. Further, $G_\sigma g_n = g_n * g_\sigma = g_{n+\sigma}$, and thus,
		\begin{IEEEeqnarray*}{c}
			\frac{\normLebTwo{\reals^2}{G_\sigma g_n}}{\normLebTwo{\reals^2}{g_n}} = \frac{n}{\sigma+n} \rightarrow 1 \mbox{ when }n\rightarrow +\infty\,.
		\end{IEEEeqnarray*}
		Therefore, \hyperref[eq:normGauss:first]{(\ref*{eq:normGauss})} is tight and $\left\|G_\sigma\right\|_{\opers{\LebTwo{\reals^2}}{\LebTwo{\reals^2}}}=1$.
	\end{IEEEproof}

	\begin{property}[Self-Adjointness of the Gaussian blur operator] \label{prop:SelfGaus}
		The Gaussian blur operators in Definition~\ref{def:GaussBlur} are self-adjoint in $\LebTwo{\reals^2}$, i.e.,
		$G_\sigma^* = G_\sigma, \forall \sigma > 0$.
	\end{property}
	\begin{IEEEproof}
		We will prove here that any convolutional operator from $\LebTwo{\reals^2}$ to $\LebTwo{\reals^2}$ with symmetric kernel is self-adjoint. 
		This will yield the desired result because the Gaussian kernel in Definition \ref{def:GaussianKernel} is symmetric.
		Let $G$ be such an operator with symmetric kernel $g$.
		Recall the definition of adjoint from Section \ref{ssec:Notation}.
		For any $f_1,f_2 \in \LebTwo{\reals^2}$, we have
		\begin{IEEEeqnarray*}{rl}
			\prodLebTwo{\reals^2}{G f_1}{f_2} & = \int_{\reals^2} \int_{\reals^2} g(\pos-\boldsymbol{\rho}) f_1(\boldsymbol{\rho}) \dint\boldsymbol{\rho} f_2(\pos) \dint \pos \\
			& = \int_{\reals^2} \int_{\reals^2} g(\boldsymbol{\rho}-\pos)  f_2(\pos) \dint \pos f_1(\boldsymbol{\rho}) \dint\boldsymbol{\rho} \\
			& = \prodLebTwo{\reals^2}{f_1}{G f_2}\,,
		\end{IEEEeqnarray*}
		and thus $G^*=G$.
	\end{IEEEproof}
	
	Properties \ref{prop:NormOneGauss} and \ref{prop:SelfGaus} will now be used in the following two proof.
	\begin{IEEEproof}[Proof - Lemma~\ref{lem:NormDiff} (Boundedness of the diffusion operator)]
			Recall that the norm of the operator $A$ is defined as
			\begin{IEEEeqnarray*}{c}
				\|A\|_{\opers{\ASpace}{\DenSpace}} = \sup_{a \in \ASpace}\left\lbrace 
					\frac{\normDen{Aa}}{\normA{a}}
				\right\rbrace\,.
			\end{IEEEeqnarray*}
			Consider, then, that 
			\begin{IEEEeqnarray}{rl}
				\IEEEyesnumber \label{eq:ProofNormDiff:Masks} \IEEEyessubnumber* \label{eq:ProofNormDiff:Masks:l1}
				\normDen{Aa}^2 
				& = \normLebTwo{\reals^2}{ w \int_{0}^{\sigmax} G_\sigma a_\sigma \dint\sigma }^2 \\
				& \leq \wninf^2 \normLebTwo{\reals^2}{ \int_{0}^{\sigmax} G_\sigma a_\sigma \dint\sigma }^2 \\
				\IEEEyesnumber \label{eq:ProofNormDiff:Jensen} \IEEEyessubnumber* \label{eq:ProofNormDiff:Jensen:l1}
				& = \eta^{-1}  \sigmax \int_{\reals^2} \left[ 
												\frac{1}{\sigmax} \int_0^{\sigmax} G_\sigma a_\sigma \dint\sigma
												\right]^2 \!\!\!\! \dint\pos \\
				& \leq \eta^{-1}  \int_{\reals^2} \int_0^{\sigmax} 
												\left[ G_\sigma a_\sigma \right]^2 \dint\sigma \dint\pos \\
				\IEEEyesnumber \label{eq:ProofNormDiff:NormGauss} \IEEEyessubnumber* \label{eq:ProofNormDiff:NormGauss:l1} 
				& = \eta^{-1}  \int_0^{\sigmax} \normLebTwo{\reals^2}{G_\sigma a_\sigma}^2 \dint\sigma \\
				& \leq \eta^{-1}  \int_0^{\sigmax} \normLebTwo{\reals^2}{a_\sigma}^2 \dint\sigma
				 = \eta^{-1} \normA{a}^2\,.
			\end{IEEEeqnarray}
			Here, $\eta = \sigmax^{-1} \wninf^{-2}$, \hyperref[eq:ProofNormDiff:Masks:l1]{(\ref*{eq:ProofNormDiff:Masks})} 
			uses the fact that $w\in L^\infty\left(\reals^2\right)$ to bound it by its maximum value,
			\hyperref[eq:ProofNormDiff:Jensen:l1]{(\ref*{eq:ProofNormDiff:Jensen})} uses Jensen's inequality on the convex function $\alpha \in\reals \mapsto \alpha^2$,
			and \hyperref[eq:ProofNormDiff:NormGauss:l1]{(\ref*{eq:ProofNormDiff:NormGauss})} uses Property~\ref{prop:NormOneGauss} to bound the norm of $G_\sigma a_\sigma, \forall \sigma > 0$.
			Therefore, $\forall a\in\ASpace$, we have that
			\begin{IEEEeqnarray*}{c}
				\frac{\normDen{Aa}^2}{\normA{a}^2} \leq \sigmax \wninf^2\,,
			\end{IEEEeqnarray*}
			and, thus,
				$\|A\|_{\opers{\ASpace}{\DenSpace}} \leq \sqrt{ \sigmax } \wninf$.
	\end{IEEEproof}

	\begin{IEEEproof}[Proof - Lemma~\ref{lem:AdjDiff} (Adjoint to the diffusion operator)]
			Recall the definition of adjoint from Section \ref{ssec:Notation}.
			For any $a\in \ASpace$, $d\in \DenSpace$, we have that 
			\begin{IEEEeqnarray}{rl}
				\IEEEyesnumber \label{eq:ProofBackDiff:Lin} \IEEEyessubnumber*  \label{eq:ProofBackDiff:Lin:l1}
				\prodDen{Aa}{d} & = \prodLebTwo{\reals^2}{\left[\int_{0}^{\sigmax} G_\sigma a_\sigma \dint\sigma\right]}{w^2 d}   \\
				& = \int_{0}^{\sigmax} \prodLebTwo{\reals^2}{G_\sigma a_\sigma}{w^2 d} \dint\sigma   \\
				& = \int_{0}^{\sigmax} \prodLebTwo{\reals^2}{a_\sigma}{ G_\sigma\left\lbrace w^2 d \right\rbrace} \dint\sigma   \\
				\IEEEyesnumber \label{eq:ProofBackDiff:ProdA} 
				& = \prodLebTwo{\reals^3}{a}{d_\mathrm{s}} = \prodA{a}{\mu d_\mathrm{s}}\,.  
			\end{IEEEeqnarray}
			Here, $d_\mathrm{s}\in\ASpace$ such that $d_\mathrm{s}=G_\sigma\left\lbrace w^2 d \right\rbrace$, 
			\hyperref[eq:ProofBackDiff:Lin:l1]{(\ref*{eq:ProofBackDiff:Lin})} uses the linearity of
			the integral and the inner product, and \eqref{eq:ProofBackDiff:ProdA} uses that because 
			$a\in\ASpace$, $a(\pos,\sigma) d_\mathrm{s}(\pos,\sigma) = 0, \forall \pos \not \in \supp{\mu}$.
			Therefore, $A^*d = \mu d_\mathrm{s}$.			
	\end{IEEEproof}
	
	We proceed by proving Lemma~\ref{lem:NullSpace}, which characterizes the nullspace of the diffusion operator in a simple but insightful way.
	
	\begin{IEEEproof}[Proof - Lemma~\ref{lem:NullSpace} (Nullspace of the diffusion operator)]
	  Let $Aa=0$. Then, 
	  \begin{IEEEeqnarray*}{rl}
	    \int_{\reals^2} (A a)(\pos)\dint\pos & = \int_{\reals^2} \int_0^{\sigmax} (G_\sigma a_\sigma)(\pos)\, \dint\sigma \dint\pos \\
	    & = \int_{\reals^2} \int_0^{\sigmax} \int_{\reals^2} g_\sigma(\pos-\boldsymbol{\rho}) a(\boldsymbol{\rho},\sigma) \,\dint\boldsymbol{\rho} \dint\sigma \dint\pos \\
	    & = \int_0^{\sigmax} \int_{\reals^2} \int_{\reals^2} g_\sigma(\pos-\boldsymbol{\rho}) \dint\pos\, a(\boldsymbol{\rho},\sigma) \,\dint\boldsymbol{\rho} \dint\sigma \\
	    & = \int_0^{\sigmax} \int_{\reals^2} a(\boldsymbol{\rho},\sigma)\, \dint\boldsymbol{\rho} \dint\sigma \\
	    & = \normLebOne{\Omega}{a_+} - \normLebOne{\Omega}{a_-} = 0\,.
	  \end{IEEEeqnarray*}
	\end{IEEEproof}

	We finalize this appendix by proving Lemma~\ref{lem:existence}, which guarantees the existence of a minimizer of the non-negative group-sparsity regularized inverse diffusion problem under discretization.
	
	\begin{IEEEproof}[Proof - Lemma~\ref{lem:existence} (Existence of a solution to the discretized problem)]
		We will prove that $\forall \epsilon > 0$, $\exists\delta>0$ such that $C(\tilde{a}) \leq \epsilon$ implies 
		$\left\|\tilde{a}\right\|_2\leq \delta$, i.e., $\tilde{a}\in\bar{\mathrm{B}}_{\delta} = \lbrace \tilde{a}\in\matrices{M,N,K} :  \left\|\tilde{a}\right\|_2 \leq \delta\rbrace$. As a consequence, we will have shown that \eqref{eq:optidisc} is equivalent to
		\begin{IEEEeqnarray}{c}\label{eq:equivalent_bounded_problem}
			\min_{\tilde{a}\in \left[\bar{\mathrm{B}}_{\delta}\right]_+} \left\lbrace \left\| \tilde{A}\tilde{a} - \dobs \right\|_{\tilde{w}}^2 + \lambda \sum_{m,n} \! \left\| \tilde{\xi} \odot \tilde{a}_{m,n} \right\|_2  \right\rbrace\,.
		\end{IEEEeqnarray}
		Because \eqref{eq:equivalent_bounded_problem} is a minimization problem of a continuous function on a closed bounded set, the extreme-value theorem guarantees that 
		it has a minimizer, and thus, \eqref{eq:optidisc} has a minimizer too.

		Consider first the simpler case $\tilde{\xi}_k>0$ for any $k\in\listint{K}$ and $\lambda>0$. 
		Then, for any $\epsilon>0$, we have that $C(\tilde{a})\leq\epsilon$ implies $f(\tilde{a})\leq\epsilon$, which, 
		for any $(m,n)$ implies that $\left\| \tilde{\xi} \odot \tilde{a}_{m,n} \right\|_2\leq \epsilon/\lambda$.
		Then, $\left\| \tilde{a}_{m,n} \right\|_2\leq \epsilon/(\lambda\min_k \tilde{\xi}_k)$ for any $(m,n)$ and
		\begin{IEEEeqnarray*}{c}
			\left\|\tilde{a}\right\|_2 =\sqrt{\sum_{m,n} \left\| \tilde{a}_{m,n} \right\|^2_2} \leq \sqrt{MN} \frac{\epsilon}{\lambda\min_k \tilde{\xi}_k}=\delta\,.
		\end{IEEEeqnarray*}
		
		Consider now the case in which either $\tilde{\xi}_k=0$ for some $k$ or $\lambda=0$. Consider the decomposition of $\tilde{a}$ on three unique components, i.e.,
		$\tilde{a}=\tilde{a}_{\bot} + \tilde{a}_{+} + \tilde{a}_{-} $, where $\tilde{a}_{\bot}$ is the component on the orthogonal complement to 
		the nullspace of $\tilde{A}$, while $\tilde{a}_{+}$ and $\tilde{a}_{-}$ are the non-negative and non-positive parts of the component in the nullspace of 
		$\tilde{A}$, i.e.,
		\begin{IEEEeqnarray*}{c}
			\tilde{a}_{\bot} \in\NullSpace{\tilde{A}}^\bot\mbox{ and }\tilde{a}_\mathrm{N} = \tilde{a}_{+} + \tilde{a}_{-}\in\NullSpace{\tilde{A}}\,.
		\end{IEEEeqnarray*}
		Then, we have that if $\epsilon\geq C(\tilde{a})$, then
		\begin{IEEEeqnarray*}{ll}
			\sqrt{\epsilon} \,\,& \geq \left\| \tilde{A}\tilde{a} - \dobs \right\|_{\tilde{w}} 
			\geq \left\| \tilde{A}\tilde{a}\right\|_{\tilde{w}} - \left\|\dobs \right\|_{\tilde{w}} \\
			& = \left\| \tilde{w} \odot \tilde{A}\tilde{a}\right\|_{2} - \left\|\dobs \right\|_{\tilde{w}} \\
			& \geq \min_{m,n}\tilde{w}_{m,n} \left\| \tilde{A}\tilde{a}\right\|_{2} - \left\|\dobs \right\|_{\tilde{w}} \\
			& \geq \kappa \min_{m,n}\tilde{w}_{m,n} \left\|\tilde{a}_{\bot} \right\|_2 - \left\|\dobs \right\|_{\tilde{w}} \,,
		\end{IEEEeqnarray*}
		where $\kappa$ is the smallest non-zero singular value of $\tilde{A}$, and, in conclusion,
		\begin{IEEEeqnarray*}{c}
			\left\|\tilde{a}_{\bot} \right\|_2 \leq
			\frac{\sqrt{\epsilon} + \left\|\dobs \right\|_{\tilde{w}}}{ \kappa \min_{m,n}\tilde{w}_{m,n}} = \delta_1\,.
		\end{IEEEeqnarray*}
		Moreover, because $C(\tilde{a})<+\infty$, the non-negative constraint must be satisfied, and therefore, 
		$\left\|\tilde{a}_-\right\|_\infty\leq\left\|\tilde{a}_\bot\right\|_\infty$. Further, because $\matrices{M,N,K}$ is a finite-dimensional space,
		for any $p,q\in[1,\infty]$, $\exists c_{p,q}\geq 0$ such that for any $\tilde{a}\in\matrices{M,N,K}$,
		$\left\|\tilde{a}\right\|_p \leq c_{p,q} \left\| \tilde{a} \right\|_q$. Then, Lemma~\ref{lem:NullSpace} yields that
 		\begin{IEEEeqnarray*}{rl}
			\left\| \tilde{a}_\mathrm{N} \right\|_2 & \leq c_{2,1} \left\| \tilde{a}_\mathrm{N} \right\|_1 = 2\, c_{2,1}  \left\| \tilde{a}_- \right\|_1 \\
			& \leq 2\, c_{2,1} c_{1,\infty} \left\| \tilde{a}_- \right\|_\infty 
			\leq 2\, c_{2,1} c_{1,\infty} \left\| \tilde{a}_\bot \right\|_\infty \\ & \leq 2\, c_{2,1} c_{1,\infty} c_{\infty,2} \left\| \tilde{a}_\bot \right\|_2\,,
 		\end{IEEEeqnarray*}
 		and 
 		\begin{IEEEeqnarray*}{rl}
			\left\| \tilde{a} \right\| & = \sqrt{\left\| \tilde{a}_\mathrm{N} \right\|^2_2 + \left\|\tilde{a}_{\bot} \right\|^2_2} \\
			& \leq \delta_1\, \sqrt{1+4 [c_{2,1} c_{1,\infty} c_{\infty,2}]^2}  = \delta\,.
 		\end{IEEEeqnarray*}
	\end{IEEEproof}
	
	\ifbool{tot}{}{
		\bibliographystyle{IEEEtran}
		\bibliography{IEEEabrv,\bib/multi_deconv}
	}

		\section*{Acknowledgments}
			
			Doctor Holger Kohr, Axel Ringh and Assistant Professor Johan Karlsson provided valuable advice on
			operator discretizations. Professon Lars Jonsson provided valuable insights and helpful discussions
			on the solutions of \eqref{eq:pde}. Doctor Christian Smedman provided expert labeling of the real Fluorospot 
			data provided by Mabtech AB.
			The excellent team of anonymous reviewers provided feedback 
			that improved the presentation of our results considerably. Of particular relevance were their pointing
			to the question of the existence of a solution to \eqref{eq:InvDif:Regularised} and their suggestion of simpler proof techniques for Property~\ref{prop:NormOneGauss}.
		

		\IEEEtriggeratref{56}

		\bibliographystyle{IEEEtranMod}
		\bibliography{IEEEabrv,\bib/multi_deconv}

		\vspace{-200pt}
		
		\begin{IEEEbiography}[{\includegraphics[width=1in,height=1.25in,clip,keepaspectratio]{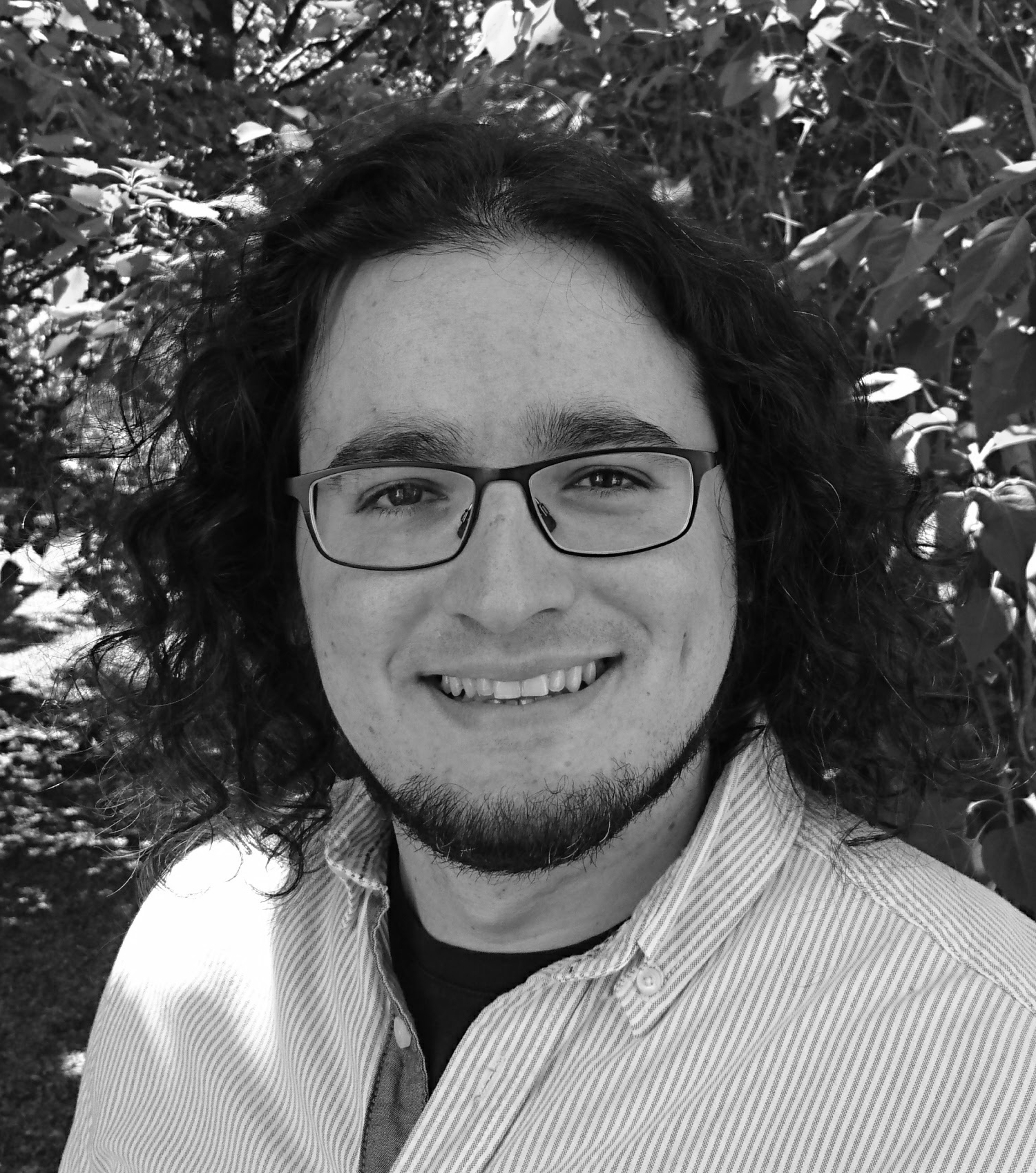}}]{Pol del Aguila Pla}
			a (S’15) received a double degree in
telecommunications and electrical engineering from the Universitat Polit\`ecnica de Catalunya, Barcelona,
Spain, and the Royal Institute of Technology (KTH),
Stockholm, Sweden, in 2014. Since August 2014,
he is currently working toward the Ph.D. degree
in electrical engineering and signal processing with
KTH under the supervision of Joakim Jalden. His
Ph.D. work includes the research collaboration with
Mabtech AB that led to the results published here
and the development of the ELISPOT and Flourospot
reader Mabtech IRIS\textsuperscript{TM} . Since August 2015, he is a Reviewer for the \textsc{IEEE
Transactions on Signal Processing}. During 2017, he received a number of
grants to support the international promotion of his research in inverse problems
for scientific imaging, including a 2017 KTH Opportunities Fund scholarship,
a Knut and Alice Wallenberg Jubilee appropriation, an Aforsk Foundation’s
scholarship for travel and a 2017 Engineering Sciences grant from the Swedish
Academy of Sciences (KVA, ES2017-0011).
		\end{IEEEbiography}		
		
		\vspace{-195pt}
		
		\begin{IEEEbiography}[{\includegraphics[width=1in,height=1.25in,clip,keepaspectratio]{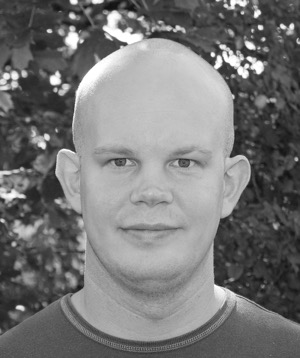}}]{Joakim Jald\'{e}n} 
			(S’03–M’08–SM’13) received the
M.Sc. and Ph.D. degrees in electrical engineering
from the Royal Institute of Technology (KTH),
Stockholm, Sweden, in 2002 and 2007, respectively.
From July 2007 to June 2009, he held a postdoctoral
research position with the Vienna University
of Technology, Vienna, Austria. He also studied at
Stanford University, Stanford, CA, USA, from
September 2000 to May 2002, and worked at ETH,
Zurich, Switzerland, as a Visiting Researcher, from
August to September, 2008. In July 2009, he returned
to KTH, where he is currently a Professor of signal processing. His recent work
includes work on signal processing for biomedical data analysis, and the automated
tracking of (biological) cell migration and morphology in time-lapse
microscopy in particular. Early work in this field was awarded a conference
best paper Award at IEEE ISBI 2012, and subsequent work by the group has
been awarded several Bitplane Awards in connection to the ISBI cell tracking
challenges between 2013 and 2015. He was an Associate Editor for the \textsc{IEEE
Communications Letters} between 2009 and 2011, and an Associate Editor
for the \textsc{IEEE
Transactions on Signal Processing} between 2012 and 2016.
Since 2013, he has been a member of the IEEE Signal Processing for Communications
and Networking Technical Committee, where he is currently a
Vice-Chair. Since 2016, he has also been responsible for the five year B.Sc and
M.Sc. Degree Program in electrical engineering with KTH.

For his work on MIMO communications, he has been awarded the IEEE
Signal Processing Societies 2006 Young Author Best Paper Award, the Distinguished
Achievement Award of NEWCOM++ Network of Excellence in
Telecommunications 2007–2011, and the best student conference paper Award
at IEEE ICASSP 2007. He is also the recipient of the Ingvar Carlsson Career
Award issued in 2009 by the Swedish Foundation for Strategic Research.
		\end{IEEEbiography}

\end{document}